\documentclass[a4paper,fleqn,usenatbib]{mnras}
\pdfoutput=1
\usepackage{newtxtext,newtxmath}
\usepackage{mathptmx}   
\usepackage{txfonts}

\usepackage[T1]{fontenc}
\usepackage{ae,aecompl}

\usepackage{graphicx}	
\usepackage{epstopdf}
\usepackage{psfig} 
\usepackage{amsmath}	
\usepackage{amssymb}	
\usepackage[T1]{fontenc}
\usepackage{aecompl}
\usepackage{multirow}
\usepackage{subcaption}	
\title[How do binary clusters form?]{How do binary clusters form?}

\author[Becky Arnold et al.]{
Becky Arnold,\thanks{E-mail: rjarnold1@sheffield.ac.uk}
Simon P. Goodwin,
D. W. Griffiths
and Richard. J. Parker\thanks{Royal Society Dorothy Hodgkin Fellow}
\\
Department of Physics and Astronomy, University of Sheffield, Sheffield S3 7RH, UK\\
}

\date{Accepted XXX. Received YYY; in original form ZZZ}

\pubyear{2017}

\begin{document}
\label{firstpage}
\pagerange{\pageref{firstpage}--\pageref{lastpage}}
\maketitle

\begin{abstract}
  
  Approximately 10 per cent of star clusters are found in pairs, known as binary
  clusters. We propose a mechanism for binary cluster formation; we
    use N-body simulations to show that velocity substructure in a single
    (even fairly smooth) region can cause binary clusters to form. This
 process is highly stochastic and it is not obvious from a region's initial conditions whether a binary will form and, if it does, which stars will end up in which cluster.
    We find the probability that a region will divide
    is mainly determined by its virial ratio,
    and a virial ratio above `equilibrium'
    is generally necessary for binary formation.
    We also find that the mass ratio of the two
    clusters is strongly influenced by the initial
    degree of spatial substructure in the
  region.
  
\end{abstract}




\section{Introduction}

Star clusters are fascinating objects as they provide crucial 
tracers of the star formation, chemical, and dynamical histories 
of galaxies.  Most star clusters are thought to form in a single star
formation event and remain as coherent bound entities following this
event. (If they disperse rapidly they are not `star clusters'
under this definition.)

An interesting observation is that star clusters are quite often 
found in pairs or higher-order systems \citep{Rozhavskii76}.
Such pairs are an expected result of chance line-ups (clusters far from each
other appearing to be close due to viewing angle, (e.g. \citealt{Conrad17}).
However, once the effects of chance line-ups are accounted for a
surplus of cluster pairs is still observed, indicating that at least some
of them are related objects which are physically close to one another.
Studies of the LMC and SMC appear to
show that roughly 10 per cent of clusters are in such pairs, which are known as binary clusters
(\citealt{Pietrzynski00}).
The fraction of binary clusters in the MW has been found to be lower than this
by some studies \citep{Subramaniam95}, and about the same by
others \citep{dFM09b}.

Binary clusters are systematically younger than single clusters
e.g. half of the clusters in binaries identified by \citet{dFM09b} are
< 25 Myr old, and almost all of those are in coeval pairs (see also
\citealt{Dieball02}; \citealt{Palma16}). This is not an unusual
result; the clusters that constitute a binary are often coeval
(e.g. \citealt{Kontizas93}; \citealt{Mucciarelli12}). 
The most obvious explanation for pairs of clusters
with very similar ages is that their formation was linked in some way,
but the origins of these pairs are not understood. 

Some multiplicity may be expected as a natural consequence of
structure in molecular clouds (e.g. \citealt{Elmegreen96}). This
  paper presents an additional mechanism for binary
  cluster formation: the division of a single star
  forming region (as seen to some degree in e.g. \citealt{Goodwin04} and
  \citealt{Parker14}). 

In this paper we present a series of $N$-body simulations.  We show
that, at least for some initial conditions, binary clusters are a fairly
common outcome of the dynamical evolution of these systems.  We describe our
initial conditions in Section~2, present detailed results from a small
set of simulations in Section 3, conduct a parameter space study in
Section 4, and and conclude in Section 5.

\section{Method}

We perform purely $N$-body simulations of fractal distributions
using the \textsc{\small{}kira} integrator, which is part of
\textsc{\small{}starlab} (\citealt{PortegiesZwart99};
\citealt{PortegiesZwart01}).  Our simulations include no gas, no stellar
evolution, and no external tidal fields.  As such, they are very
simple numerical experiments, but we argue that they capture all
of the essential physics of a possible binary cluster formation mechanism.
We run the simulations for 20~Myr.

\subsection{Positions and masses}

Artificial young star forming regions are constructed using the box
fractal method, which is described in detail in  \citet{Goodwin04}.
In brief, box fractals are generated by creating a cube and placing a
`parent'\ star at its centre. The cube is divided into subcubes, which
have `child'\ stars placed at their centres, with noise added to avoid
a gridlike structure. Parent stars are deleted, and the children
become the new generation of parents. This process is repeated until
the desired number of stars, $N$, has been
overproduced. Finally, a sphere of radius $R$ is
cut from the initial box, and stars  
are randomly deleted until the $N$ stars remain.
We take regions with $N$~=~$1000$ and $R$~=~$2$~pc as our `standard'.  
The degree of
substructure (space-filling) is set by the fractal dimension
$D$ (e.g. 1.6 is very substructured, 3 is roughly uniform density).

The stars are assigned masses drawn randomly from the Maschberger IMF
\citep{Maschberger13} with the scale parameter $\mu$~=~$0.2~M_\odot$, and the high mass exponent $\alpha_{IMF}$~=~$2.3$.
The low mass exponent is calculated using $\beta$~=~$1.4$.
The lower and upper mass limits used are 0.1 and 50 $M_\odot$. The
Maschberger IMF is similar to the Chabrier IMF \citep{Chabrier03} and the Kroupa IMF \citep{Kroupa02}.

\subsection{Velocity structure}

As will become apparent later in the results, the velocity structure
of the fractal is very important.  The gas from which stars form is
known to have complex (turbulent) structure \citep{Larson81}, and
the stars in young regions appear to retain this structure
(e.g. \citealt{Furesz06}; \citealt{Jeffries14}; \citealt{Tobin15};
\citealt{Wright16}). 

To mimic this we assign stars velocities such that the regions are velocity coherent;
stars that form near each other have initially similar velocities.
We do this by one of two methods: either inheriting
velocities from their parent as the fractal is generated
(this is the main method used), or by imposing velocities from a divergence-free turbulent
velocity field.

{\bf Inherited velocities:} Following  \citet{Goodwin04},
parent stars at the first level are given a random velocity.
Child stars inheret their parent's velocity plus a random component
which scales with the depth in the fractal (i.e. the random component
is large at the higher-levels, and becomes smaller). This
creates a velocity field in which stars that are close together in
space tend to have initially similar velocities. 

Fig. 1 shows an $N$~=~$100$ fractal produced by this method. Star
positions are indicated by red dots plotted in three dimensional
space, and their velocities by arrows. As can be seen, the velocity
field has local `coherence'. For example, the stars on the
upper far-left of the figure are all moving to the right, while
on the lower far-left there is a group of four stars moving
downwards and slightly to the left.

{\bf Turbulent velocity fields:} We generate divergence-free turbulent velocity
fields with a power spectrum $P$($k$)~=~$k^{-\alpha}$ for the region,
where $\alpha$~=~2 (e.g. \citealt{Burkert00}; \citealt{Lomax14}).
  The initial positions of the stars are mapped onto these
  fields, and they are assigned the
field velocity at their location.

\subsection{Virial ratio}

Finally, the velocities are scaled to set the desired virial ratio,
$\alpha_{\rm vir}$~=~$T$~/~$|\Omega|$ (where $T$ is the total kinetic energy,
and $\Omega$ the  
total potential energy). 

\subsection{Ensembles}

We perform simulations with fractal dimensions
of $D$~=~$1.6$, 2.2 and 2.9. When $D$~=~$1.6$ we describe the
simulations as highly-substructured (`H'), $D$~=~$2.2$ is
moderately-substructured (`M'), and
smooth (`S') when $D$~=~$2.9$.
The velocities are scaled to have a virial ratio of $\alpha_{\rm vir}$~=~$0.3$
(cool, `C'), $0.5$ (virialised, `V'), or 0.7 (warm, `W').

We refer to the initial conditions of a simulation by these identifying
letters, e.g. `MW' is a moderately-substructured, warm region ($D$~=~$2.2$,
$\alpha_{\rm vir}$~=~$0.7$). A summary of the simulations is shown in Table~1.

For each set of initial conditions we run an ensemble of 50
simulations in which only the random number seed used to set the
initial conditions is changed.

\begin{figure}
  \includegraphics[width=\columnwidth]{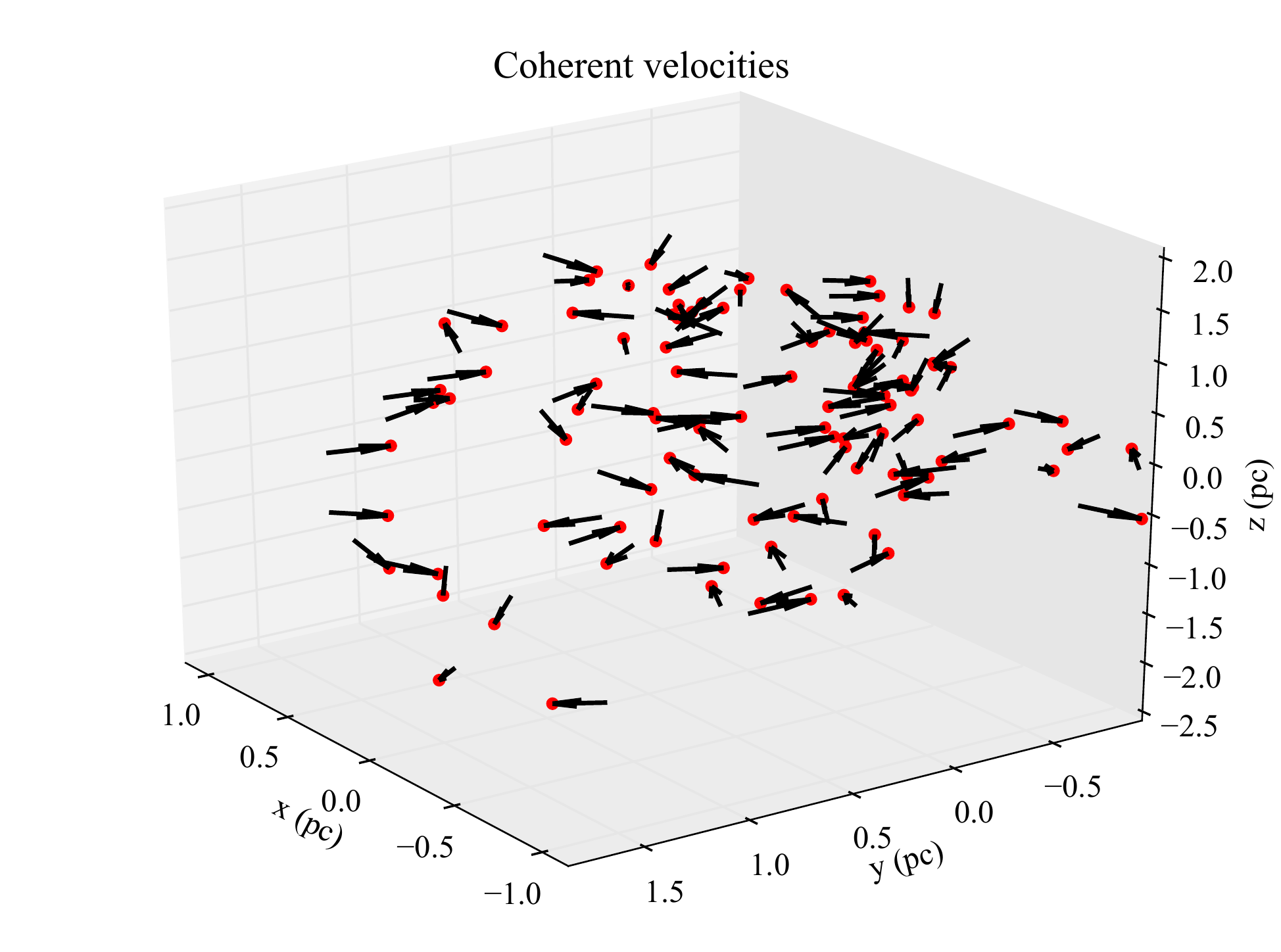}
  \caption{A set of initial conditions demonstrating velocity structure in a region with 100 stars and a radius of 1 pc. The red dots indicate the positions of stars, and their velocity vectors are denoted by black arrows.}
  \label{coherence_demo}
\end{figure}

\begin{table}
 \caption{
Letters are used to describe the initial conditions in each set of
  simulations. Two parameters are varied: the fractal dimension $D$,
  and the virial ratio (i.e. the ratio of
  kinetic to potential energy) $\alpha_{\rm vir}$.
  Highly-substructured simulations ($D$~=~$1.6$) are denoted
  by the letter
  `H', moderate substructure ($D$~=~$2.2$) is denoted by the letter `M', and
  smooth structure  ($D$~=~$2.9$) by `S'. Simulations of cool regions 
  ($\alpha_{\rm vir}$~=~$0.3$) are denoted by `C', virialised regions
  ($\alpha_{\rm vir}$~=~$0.5$) are denoted by `V', and warm regions 
($\alpha_{\rm vir}$~=~$0.7$) by `W'.  
}
  \begin{center}
\begin{tabular*}{\columnwidth}{c @{\extracolsep{\fill}} ccccc}
  \hline \\[-1.5ex]
  & & & $D$ &\\
  & & 1.6 & 2.2 & 2.9\\
  \hline \\[-1.5ex]
  \multirow{3}{1.7em}{$\alpha_{\rm vir}$}   
  & 0.3 & HC & MC & SC \\
  & 0.5 & HV & MV & SV \\ 
  & 0.7 & HW & MW & SW \\[-1.5ex] \\
  \hline
\end{tabular*}
\end{center}
\label{tab:multicol}
\end{table}

\subsection{Cluster finding}

In Appendix A we describe our cluster finding algorithm.  This is used
to distinguish bound `clusters' within our larger
regions as they evolve.  It is able to determine which stars are
locally bound to a particular object, and which are `halo' stars.
The algorithm is not perfect, and sometimes struggles when applied to
regions with ambiguous or unusual morphologies.
However it allows us to avoid `by-eye'
determinations of membership when the region evolves to a distinct
single or binary cluster, which occurs in the vast majority of cases.

\section{The formation of a binary cluster}

First we will examine the process of binary cluster formation in
a small set of simulations. This allows us to investigate the process
in detail. Twelve MW simulations ($D$~=~$2.2$, $\alpha_{\rm vir}$~=~$0.7$)
are chosen at random for this; binary cluster formation
is fairly common in the MW ensemble, and the number 12 is chosen
to produce an easily readable figure.
Later we will examine parameter space to see which
initial conditions are most likely to form binary clusters.

\subsection{An ensemble of moderately-substructured, warm regions}

In Fig.~2 we show the stellar distributions of each of the 12
regions after 20~Myr. The distributions are presented in $x$-$y$
projection in 35~pc-by-35~pc boxes.
These simulations all use inherited velocities (see Section 2.2).

A visual inspection of Fig.~2 shows that distinct binary clusters form in 4 of
the 12 realisations, in subfigures
(c), (d), (g), and 
(i)\footnote{Interestingly, in (c), (d) and (g) there are `bridges' of stars
linking the two clusters.  These `bridges' are present when viewed in
3D suggesting they are real features.  Similar bridges have been found in
observations of binary clusters (\citealt{Dieball98};
\citealt{Dieball00b}; \citealt{Minniti04}).}.

Simulation (f) has an overdensity
at roughly (-2~pc, -7~pc) which could be a small
  companion cluster. Despite the other structure in the
  region and the significant halo, our cluster-finding
    algorithm does 
  distinguish it as a distinct entity. We therefore define simulation
  (f) as a binary cluster.

Simulations (a), (b), (h), (j), (k) and (l) have evolved into 
single, central star clusters.

Simulation (e) has also evolved into a single cluster, but is
elongated.
Elongated clusters are discussed 
more later, in Section 4.3.

It is important to remember that all 12 simulations had
statistically the same initial conditions, only the random
number seed has been changed. The wide range of morphologies apparent at the
end of the simulations is not particularly surprising, as the
evolution of substructured initial conditions is known to be highly stochastic
(\citealt{Parker12}; \citealt{Alison10}).

\subsection{Future evolution}

We may na\"{\i}vely expect
binary clusters to orbit one another in the same manner binary stars do,
however in these simulations the two clusters move
directly apart, and are usually unbound from each-other.
At the end of the simulations
two of the binaries are unbound, two are {\em just} unbound,
and one is bound. The bound binary could recombine
at some point in the future, in fact such recombanations are
observed in the full ensemble of simulations,
and are discussed in Section 4. In reality such mergers
may be less likely as the Galactic
potential could shear the clusters away from each other.

Typically the relative velocities of the clusters in our simulations
are only $\sim
1$~km~s\textsuperscript{-1}, so they could remain observationally
associated for many 10s of Myr even if they are formally unbound.

\begin{figure*}
  \setcounter{figure}{1}
         \begin{subfigure}[b]{0.3\textwidth}
                 \centering
                 \includegraphics[width=0.95\textwidth]{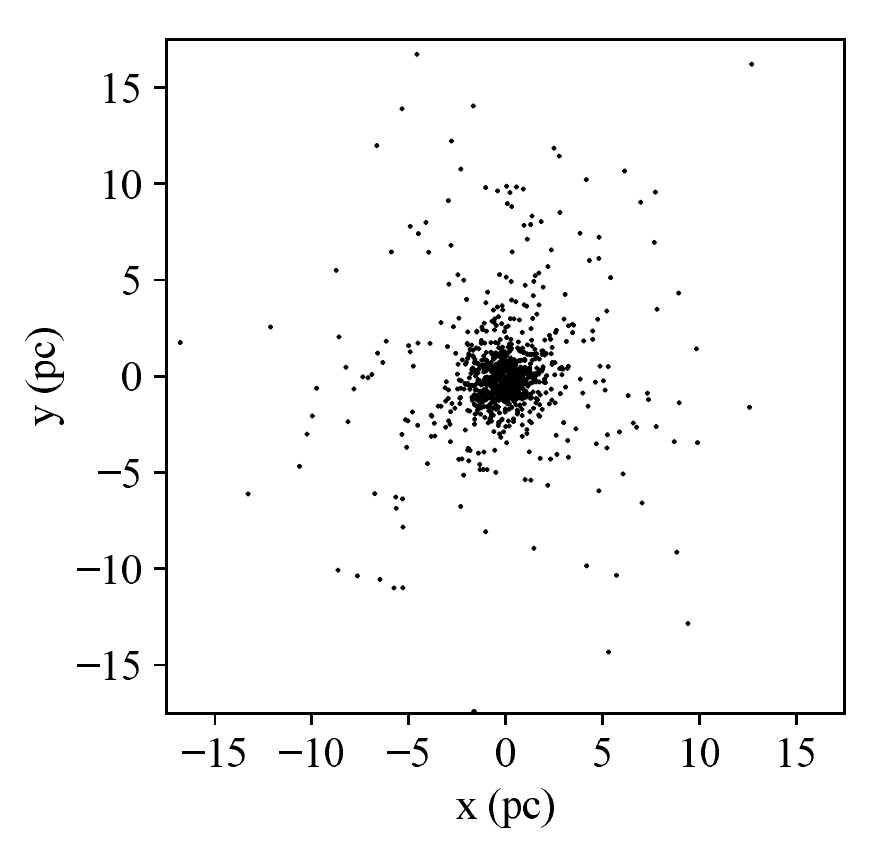}
                 \caption{a)}
                 \label{FinalStateA}
         \end{subfigure}
         \begin{subfigure}[b]{0.3\textwidth}
                 \centering
                 \includegraphics[width=0.95\textwidth]{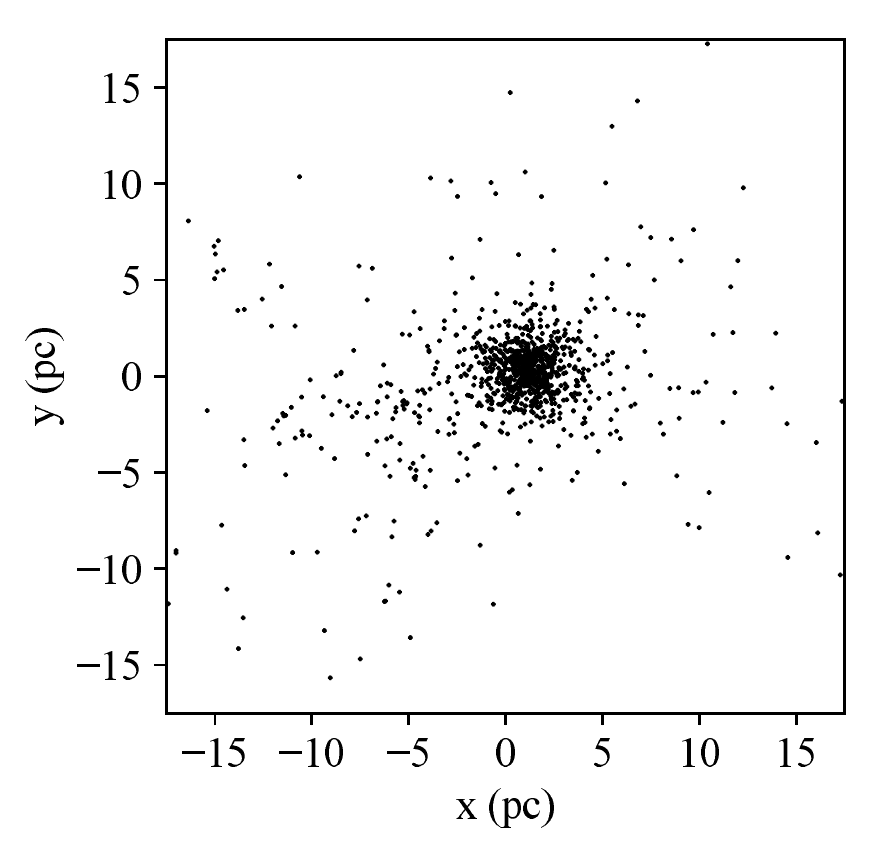}
                 \caption{b)}
                 \label{FinalStateB}
         \end{subfigure}
                  \begin{subfigure}[b]{0.3\textwidth}
                 \centering
                 \includegraphics[width=0.95\textwidth]{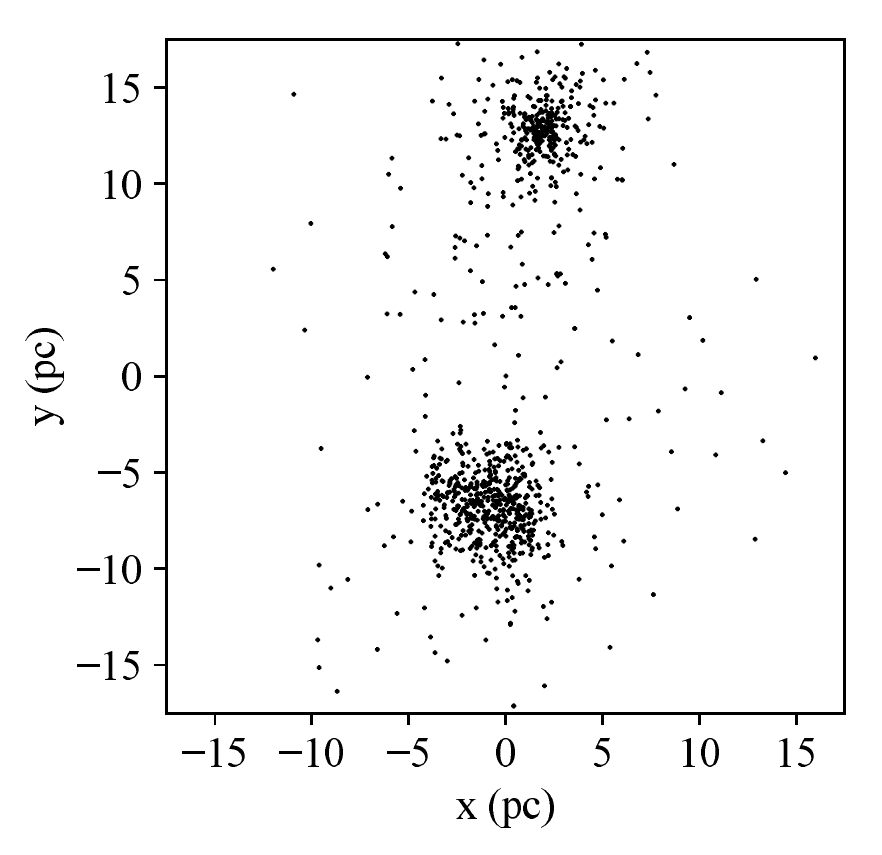}
                 \caption{c)}
                 \label{FinalStateC}
         \end{subfigure}

         \begin{subfigure}[b]{0.3\textwidth}
                 \centering
                 \includegraphics[width=0.95\textwidth]{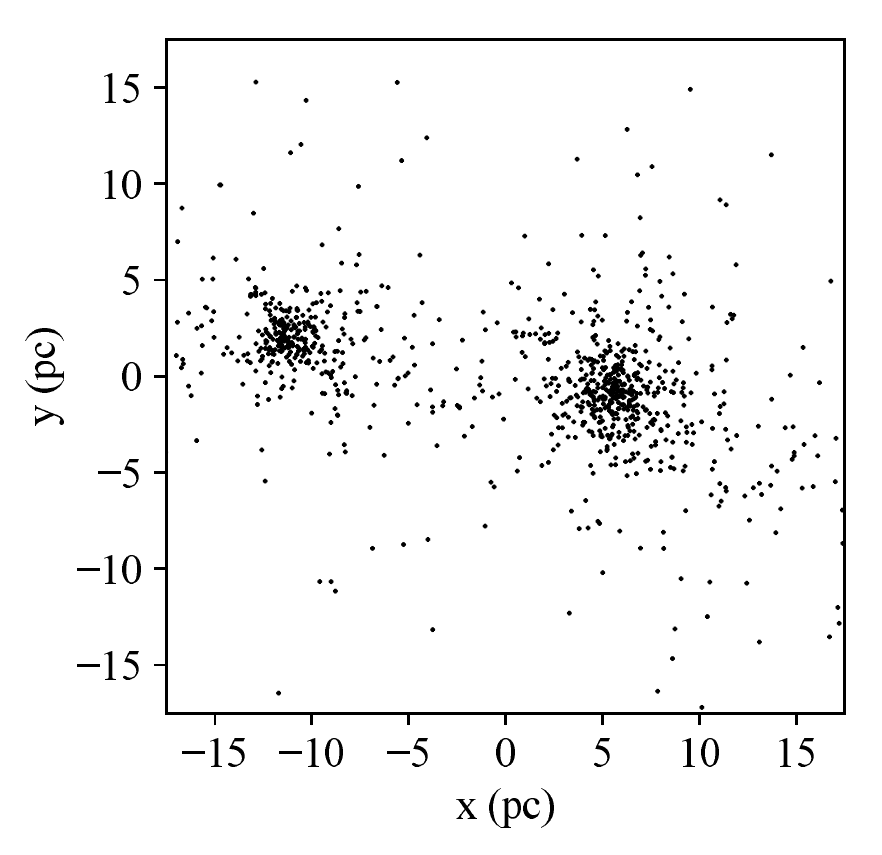}
                 \caption{d)}
                 \label{FinalStateD}
         \end{subfigure}
         \begin{subfigure}[b]{0.3\textwidth}
                 \centering
                 \includegraphics[width=0.95\textwidth]{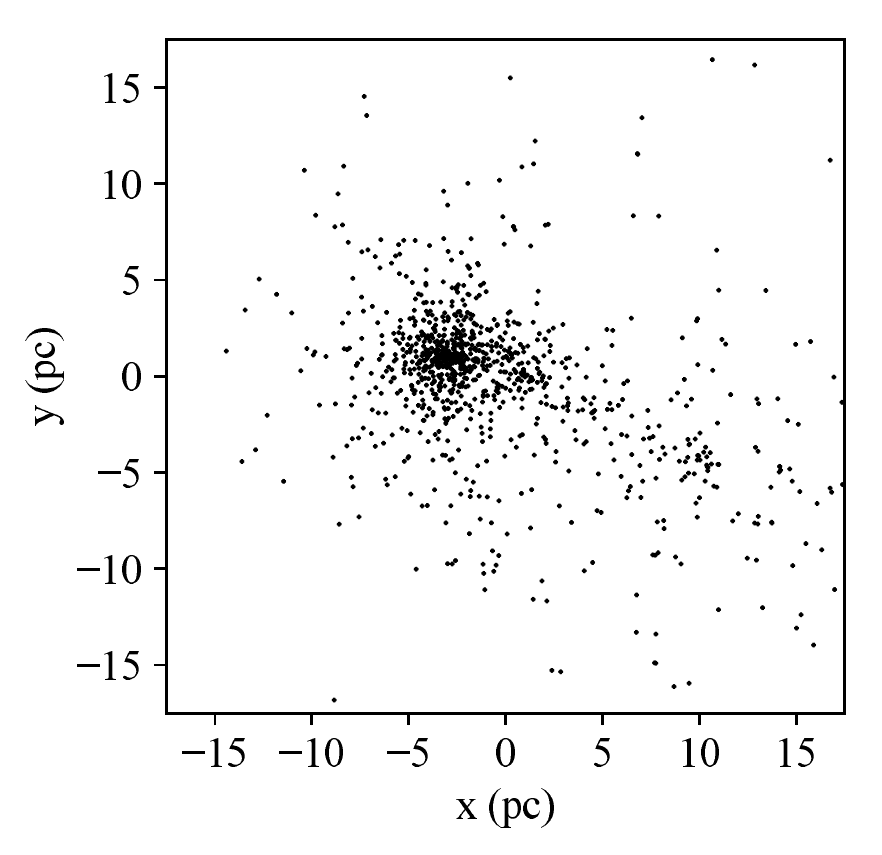}
                 \caption{e)}
                 \label{FinalStateE}
         \end{subfigure}
                  \begin{subfigure}[b]{0.3\textwidth}
                 \centering
                 \includegraphics[width=0.95\textwidth]{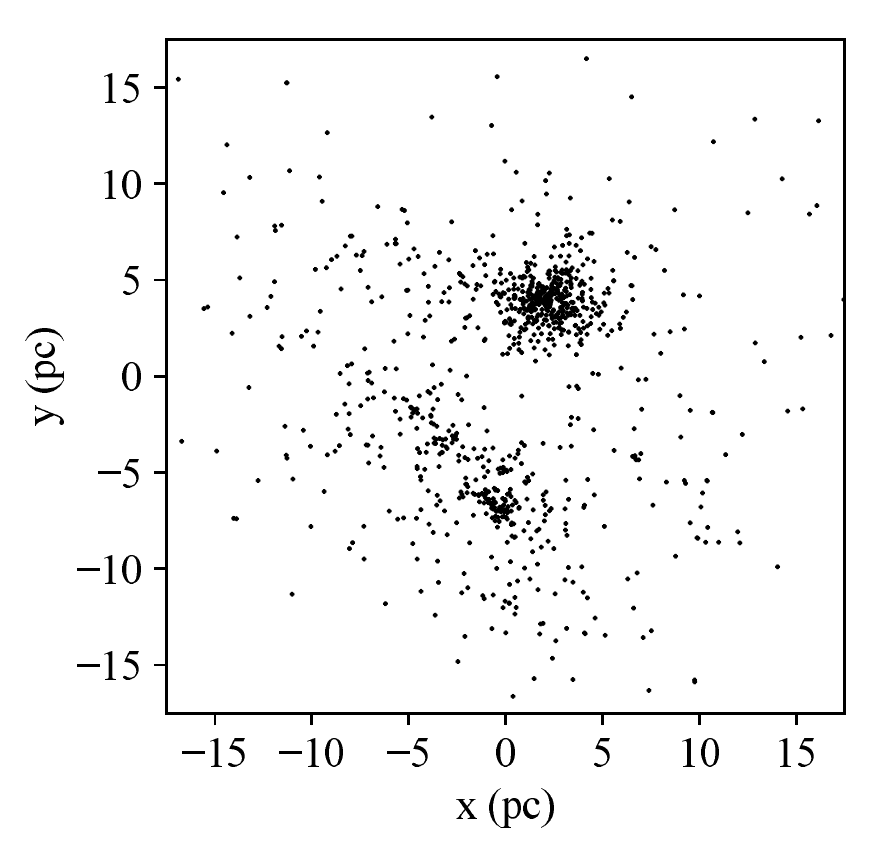}
                 \caption{f)}
                 \label{FinalStateF}
         \end{subfigure}

         \begin{subfigure}[b]{0.3\textwidth}
                 \centering
                 \includegraphics[width=0.95\textwidth]{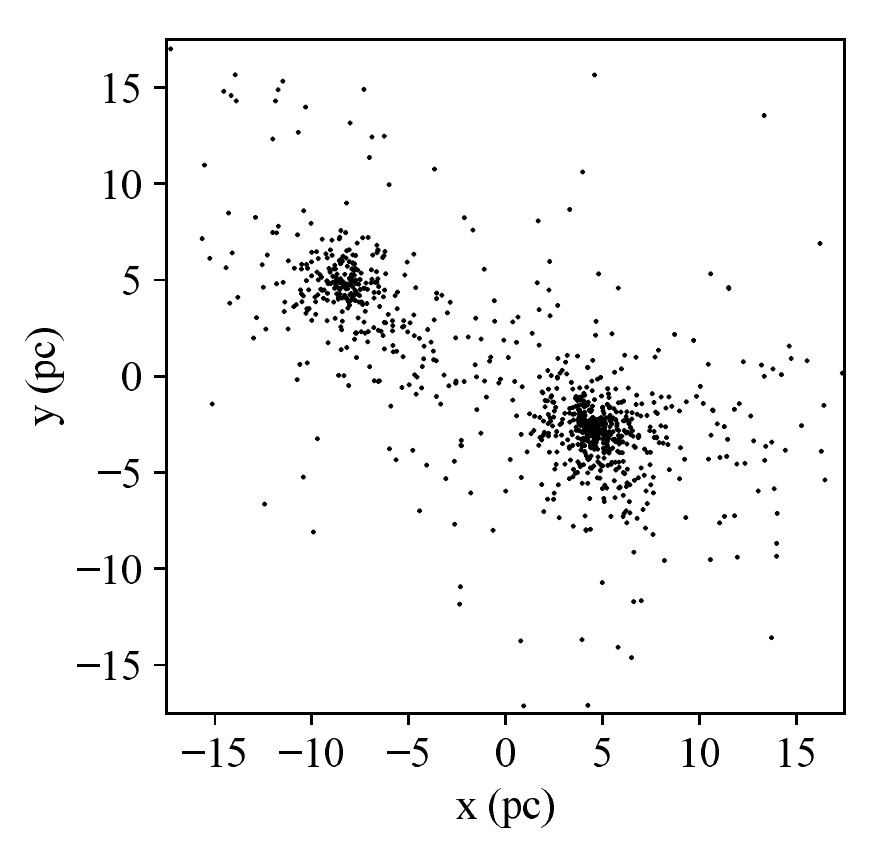}
                \caption{g)}
                 \label{FinalStateG}
         \end{subfigure}
         \begin{subfigure}[b]{0.3\textwidth}
                 \centering
                 \includegraphics[width=0.95\textwidth]{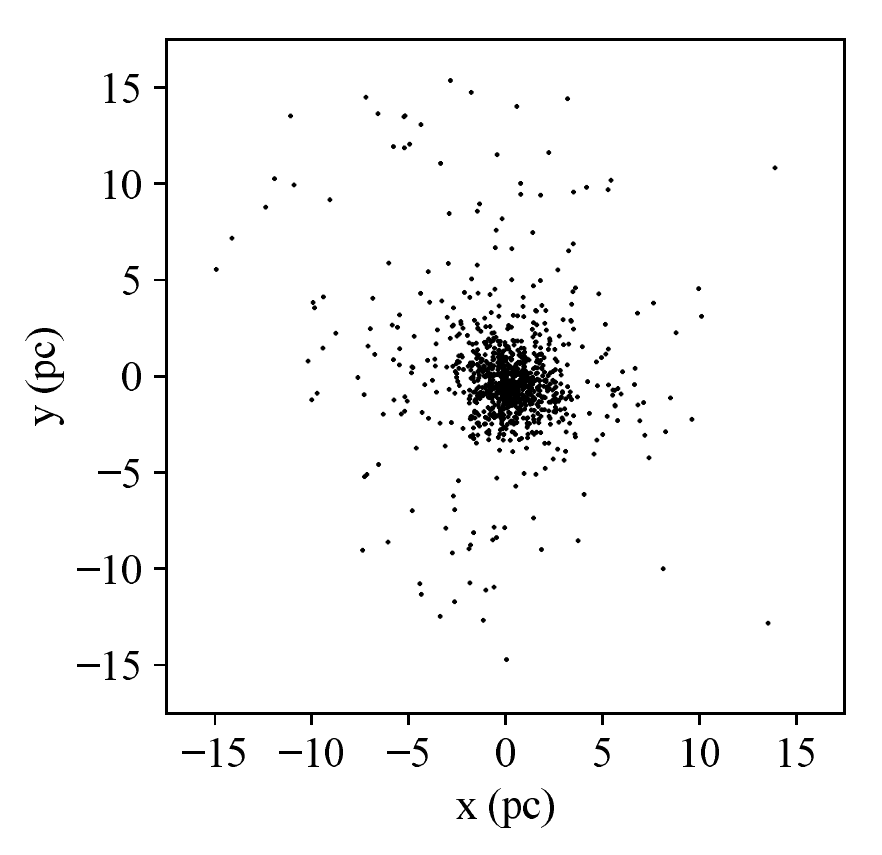}
                 \caption{h)}
                 \label{FinalStateH}
         \end{subfigure}
                  \begin{subfigure}[b]{0.3\textwidth}
                 \centering
                 \includegraphics[width=0.95\textwidth]{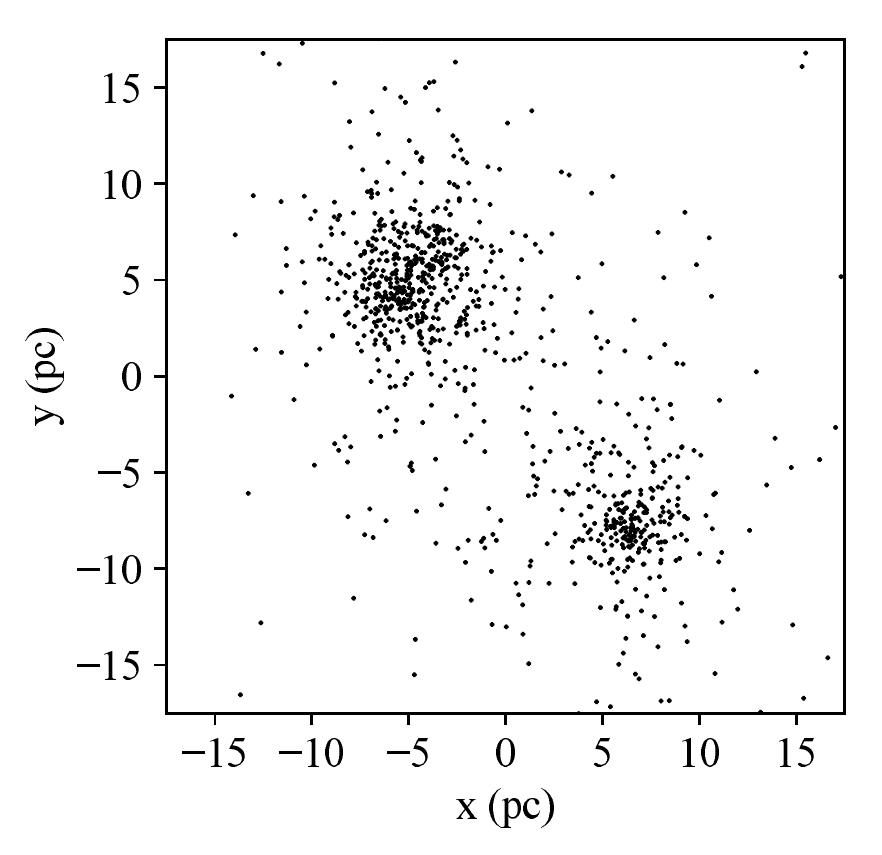}
                 \caption{i)}
                 \label{FinalStateI}
         \end{subfigure}

         \begin{subfigure}[b]{0.3\textwidth}
                 \centering
                 \includegraphics[width=0.95\textwidth]{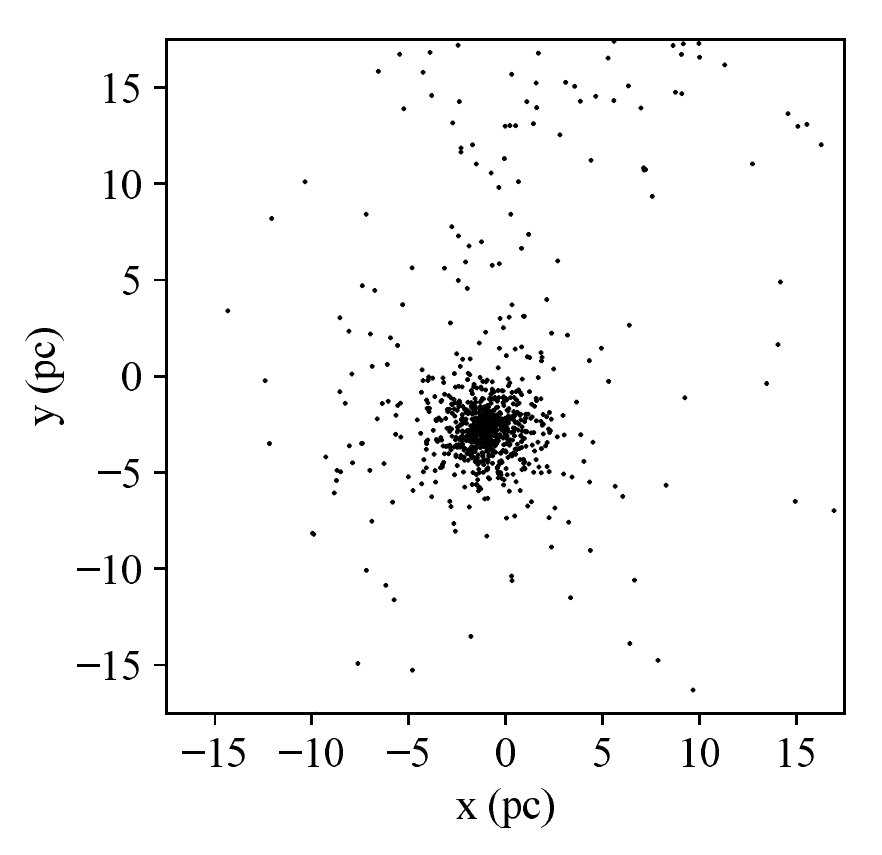}
                 \caption{j)}
                 \label{FinalStateJ}
         \end{subfigure}
         \begin{subfigure}[b]{0.3\textwidth}
                 \centering
                 \includegraphics[width=0.95\textwidth]{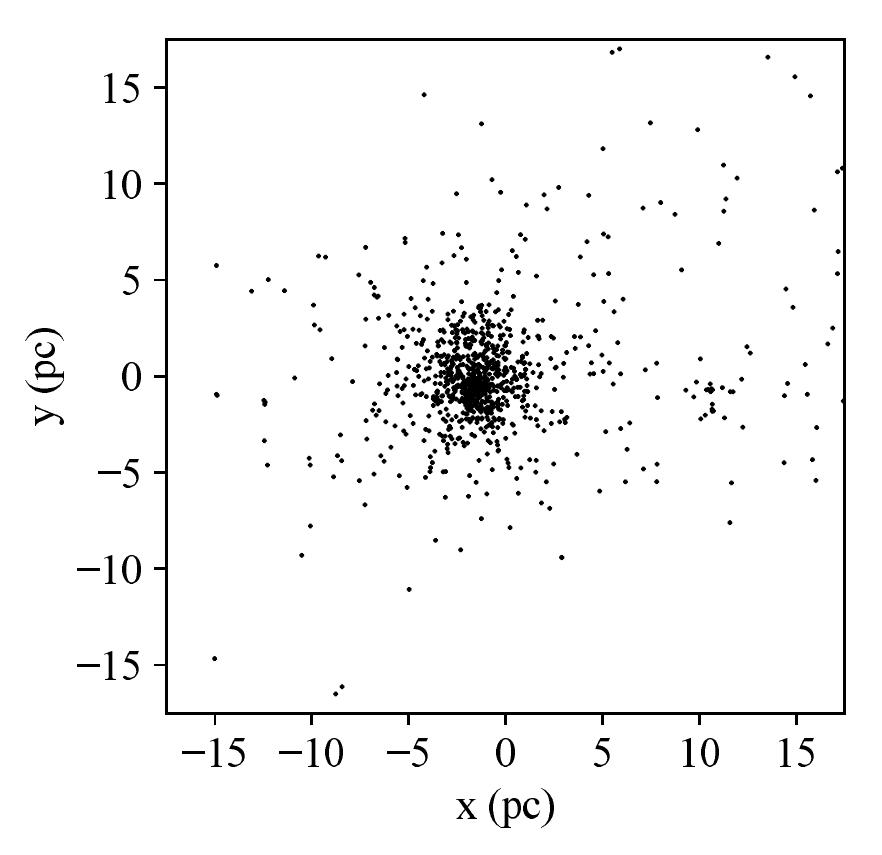}
                 \caption{k)}
                 \label{FinalStateK}
         \end{subfigure}
         \begin{subfigure}[b]{0.3\textwidth}
                 \centering
                 \includegraphics[width=0.95\textwidth]{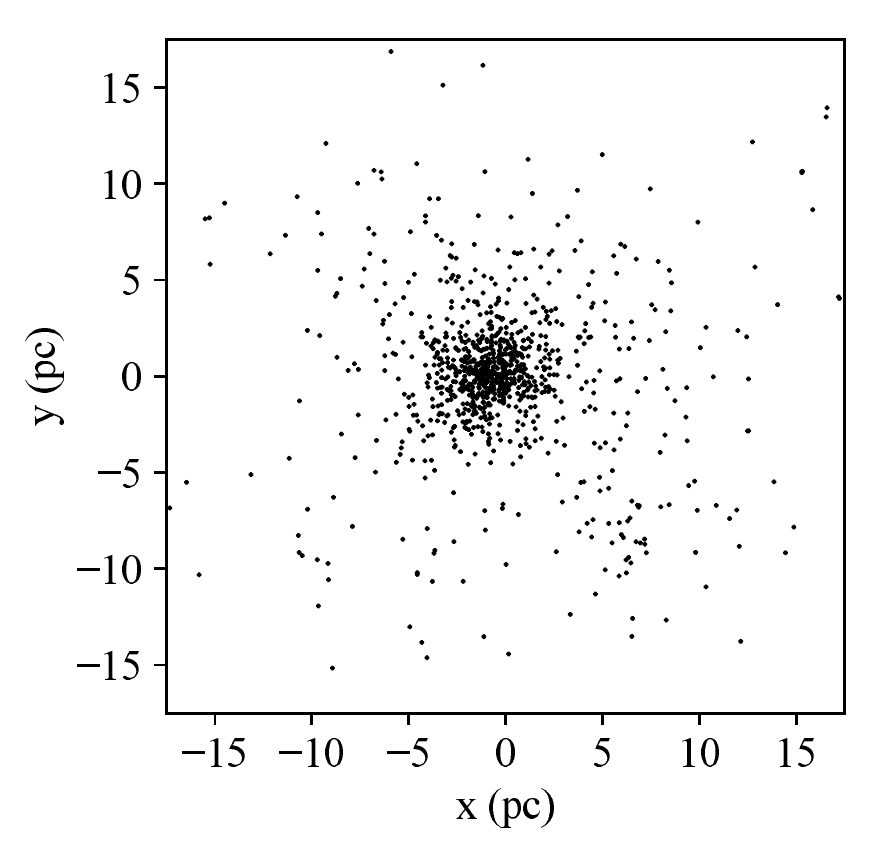}
                 \caption{l)}
                 \label{FinalStateL}
         \end{subfigure}
         \caption{35~pc-by-35~pc $x$-$y$ projections of 12 realisations of warm, moderately-substructured initial conditions that have been evolved for 20 Myr.  The only difference between realisations is the random number seed used. Every realisation contains 1000 stars.}
         \label{Final_state_plot}
\end{figure*}

\subsection{The division of a star forming region}

We now examine how binary clusters form in more detail. In Fig. 3 we
show the evolution of the region from  panel (c) of
Fig.~2 for the first 2~Myr of its evolution in steps of 0.4~Myr 
(in panel (c) of Fig.~2 the region is 20~Myr old).  
We identify the two clusters at 3~Myr when they are distinct,
well-separated entities with our cluster finding algorithm. Then, at
each time we colour code 
the stars by {\em which cluster they will eventually be members of},
blue for the cluster on the left, 
red for the cluster on the right, and black for unbound to either cluster.
Therefore, all of the red stars in the top left panel at 0~Myr
are the same stars as are coloured red in the final panel (and all
panels inbetween). For each star we also plot its velocity vector
(an arrow pointing from the position of the star).

It is immediately obvious from inspection of Fig.~3 that the stars
from each cluster are initially very well-mixed.  The red and blue stars (that
will end-up in the right- and left-hand clusters respectively) are
each found everywhere in the region at 0~Myr. Without the colour coding
(which is based on where we know they will be in the future),
from the positions of the stars alone it would 
be (a) impossible to tell that this region would evolve into a
binary cluster, and (b) impossible to tell which stars would end-up in
which cluster. This is true for {\em all} the simulations in this paper.

As the region evolves the stars which will end-up in each cluster
begin to separate out into two distinct sub-clusters.  At 0.4~Myr
there has been some separation; while the three 
classes of stars are still generally mixed, clumps of just red or
blue stars have begun to form. After 0.8 Myr there has been
further separation, and these clumps appear to have grown. By 1.2~Myr
the blue stars 
are predominantly on the left, and the red stars are predominantly on
the right. At 1.6~Myr the two groups of stars have formed roughly
spherical shapes, but it isn't until 2~Myr that clusters are well separated. 

This behaviour appears to be the result of the initial velocity
coherence.  Inspection of Fig.~3 shows that whilst the red and blue
points are initially mixed, they are not completely randomly
distributed. Even
at the very beginning there are small groups of either red
or blue stars with low velocity dispersion. These groups go on
to merge with other groups with (usually) similar
velocities.
The details of the velocity structure in this particular
case mean that a significant number of stars move in roughly the same
two directions.  In cases where the velocity structure is such that
they tend to move in many different directions then a single cluster
is formed.

\begin{figure*}
    \begin{subfigure}[b]{0.48\textwidth}
    \centering
    \includegraphics[width=1.\textwidth]{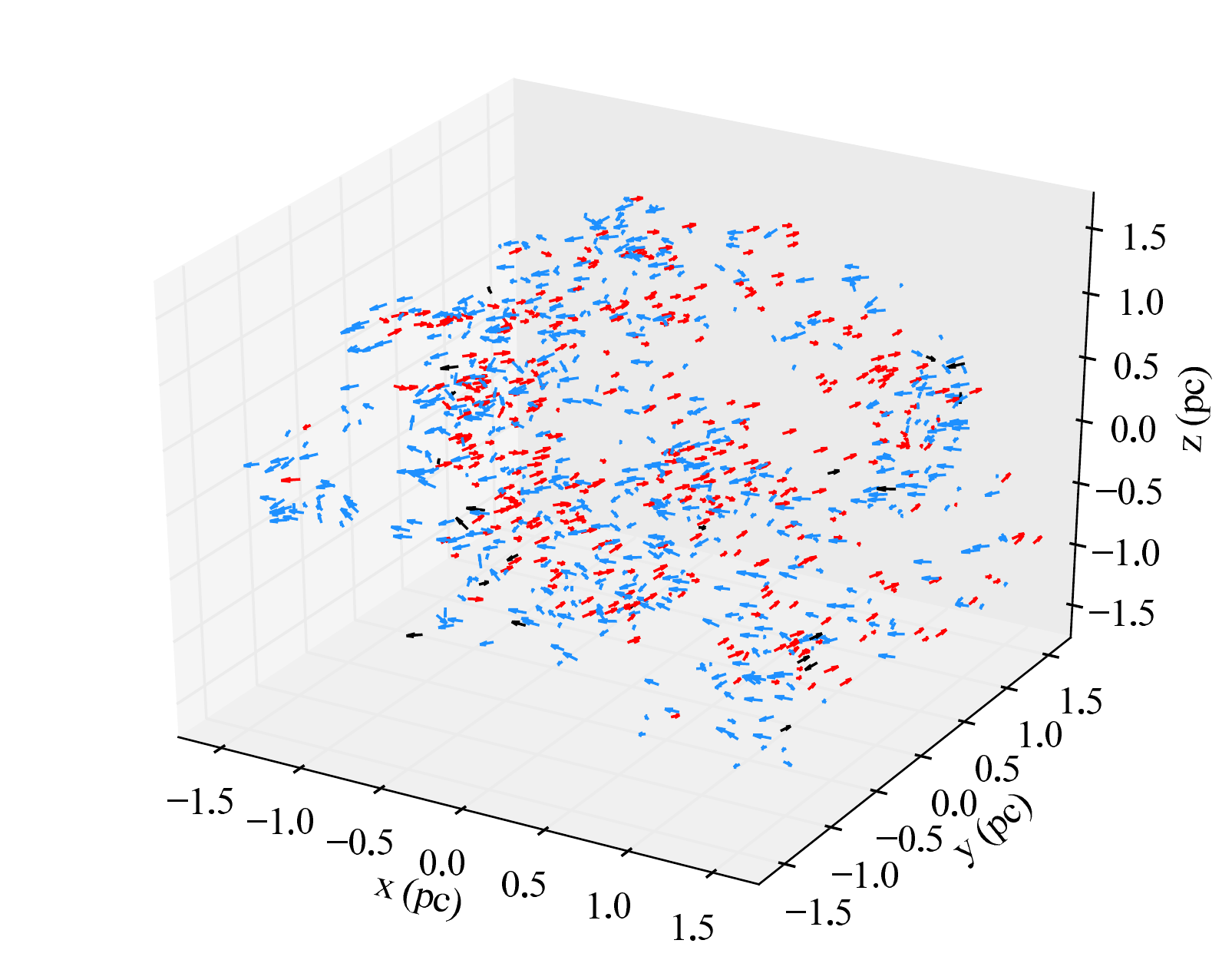}
    \caption{0 Myr}
    \label{0Myr}
  \end{subfigure}
  \begin{subfigure}[b]{0.48\textwidth}
    \centering
    \includegraphics[width=1.\textwidth]{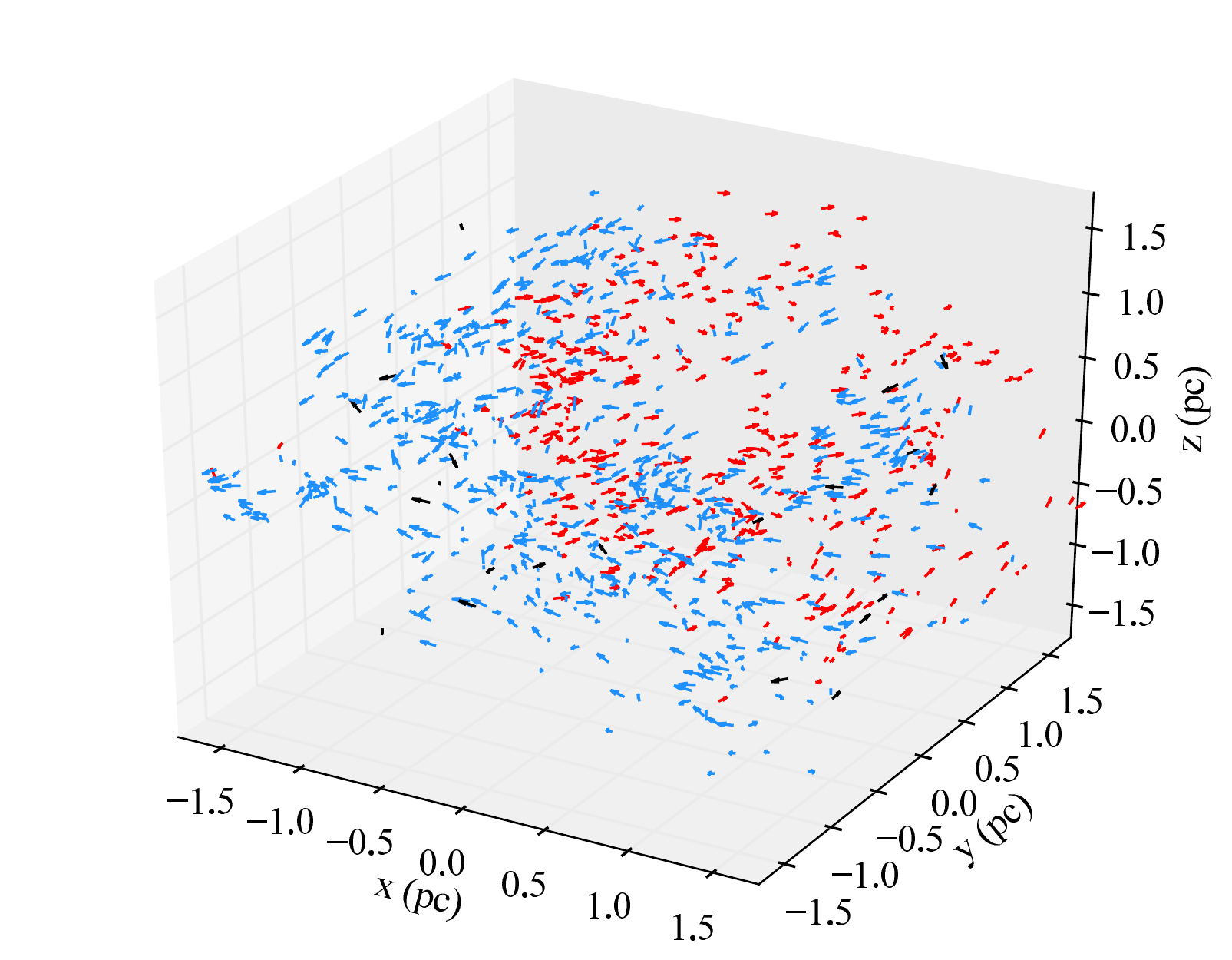}
    \caption{0.4 Myr}
    \label{0.4Myr}
  \end{subfigure}

  \begin{subfigure}[b]{0.48\textwidth}
    \centering
    \includegraphics[width=1.\textwidth]{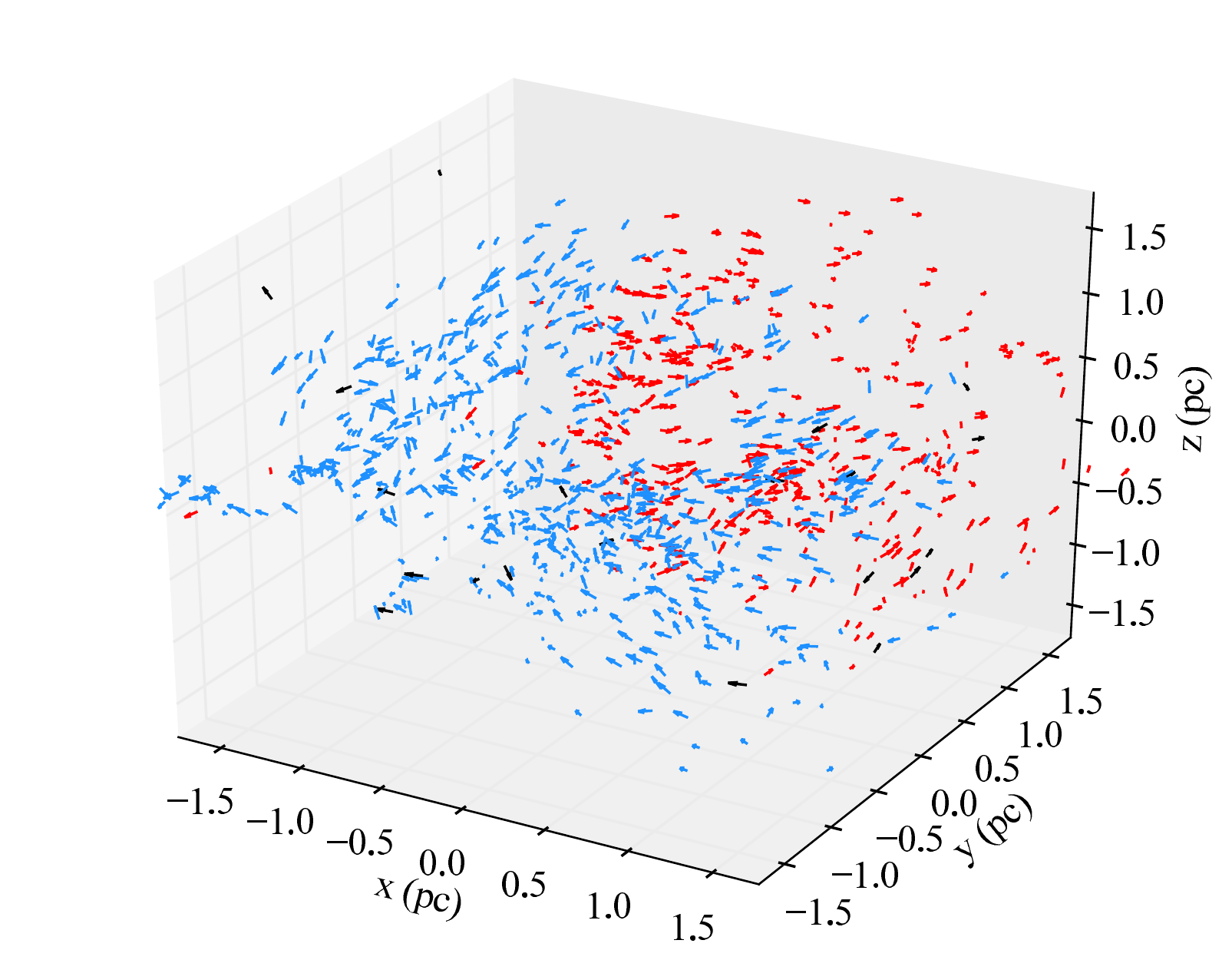}
    \caption{0.8 Myr}
    \label{0.8Myr}
  \end{subfigure}
  \begin{subfigure}[b]{0.48\textwidth}
    \centering
    \includegraphics[width=1.\textwidth]{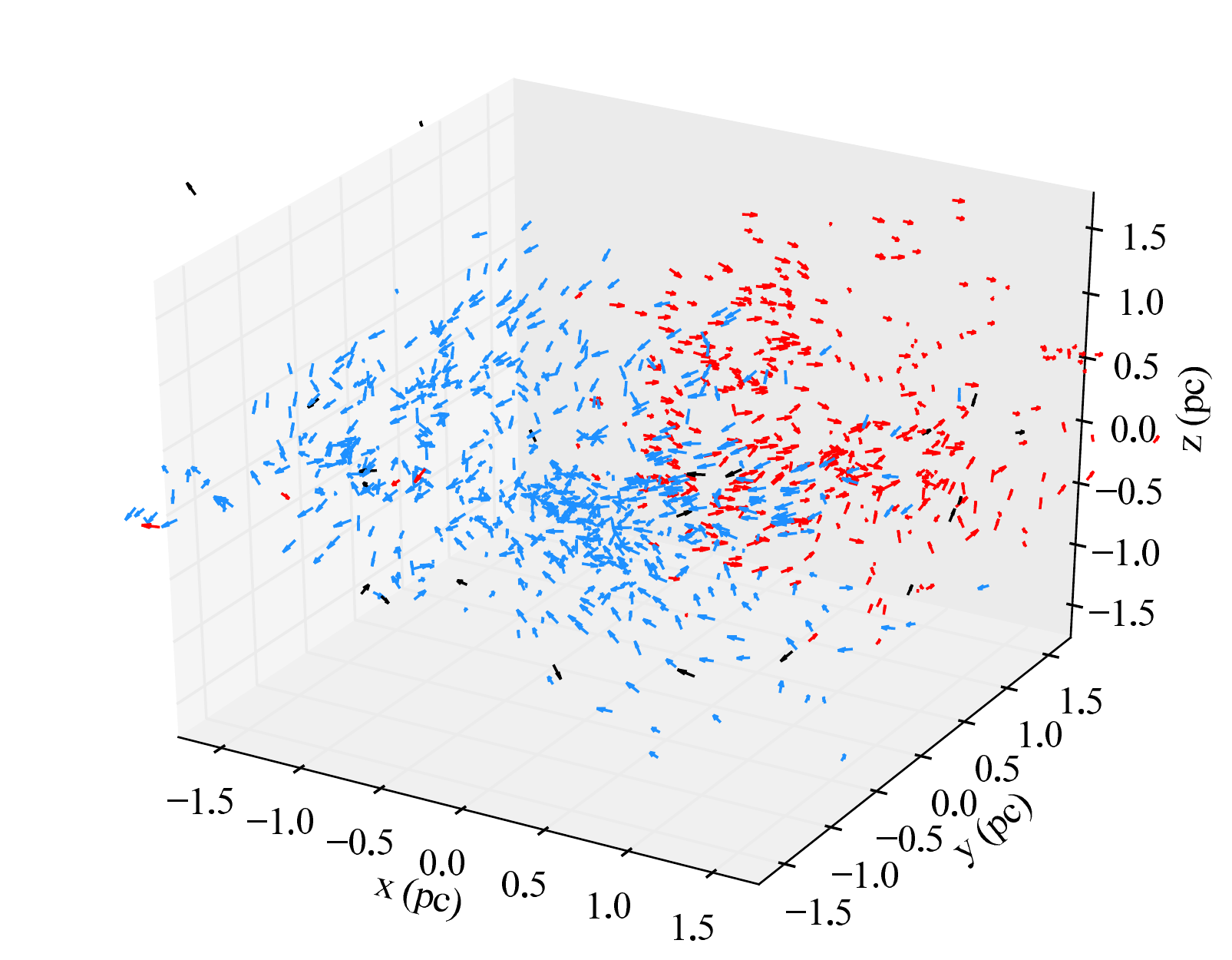}
    \caption{1.2 Myr}
    \label{1.2Myr}
  \end{subfigure}

  \begin{subfigure}[b]{0.48\textwidth}
    \centering
    \includegraphics[width=1.\textwidth]{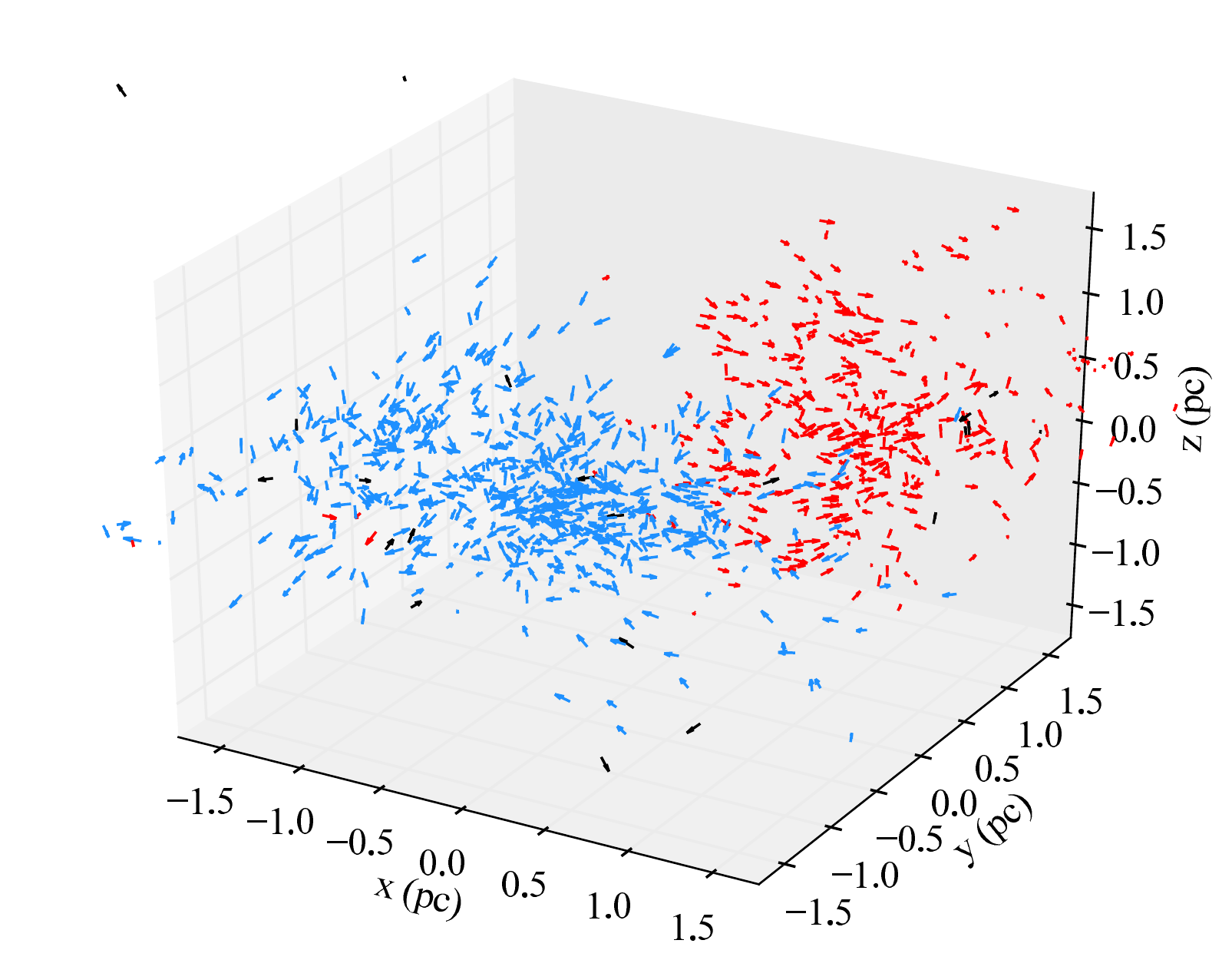}
    \caption{1.6 Myr}
    \label{1.6Myr}
  \end{subfigure}
  \begin{subfigure}[b]{0.48\textwidth}
    \centering
    \includegraphics[width=1.\textwidth]{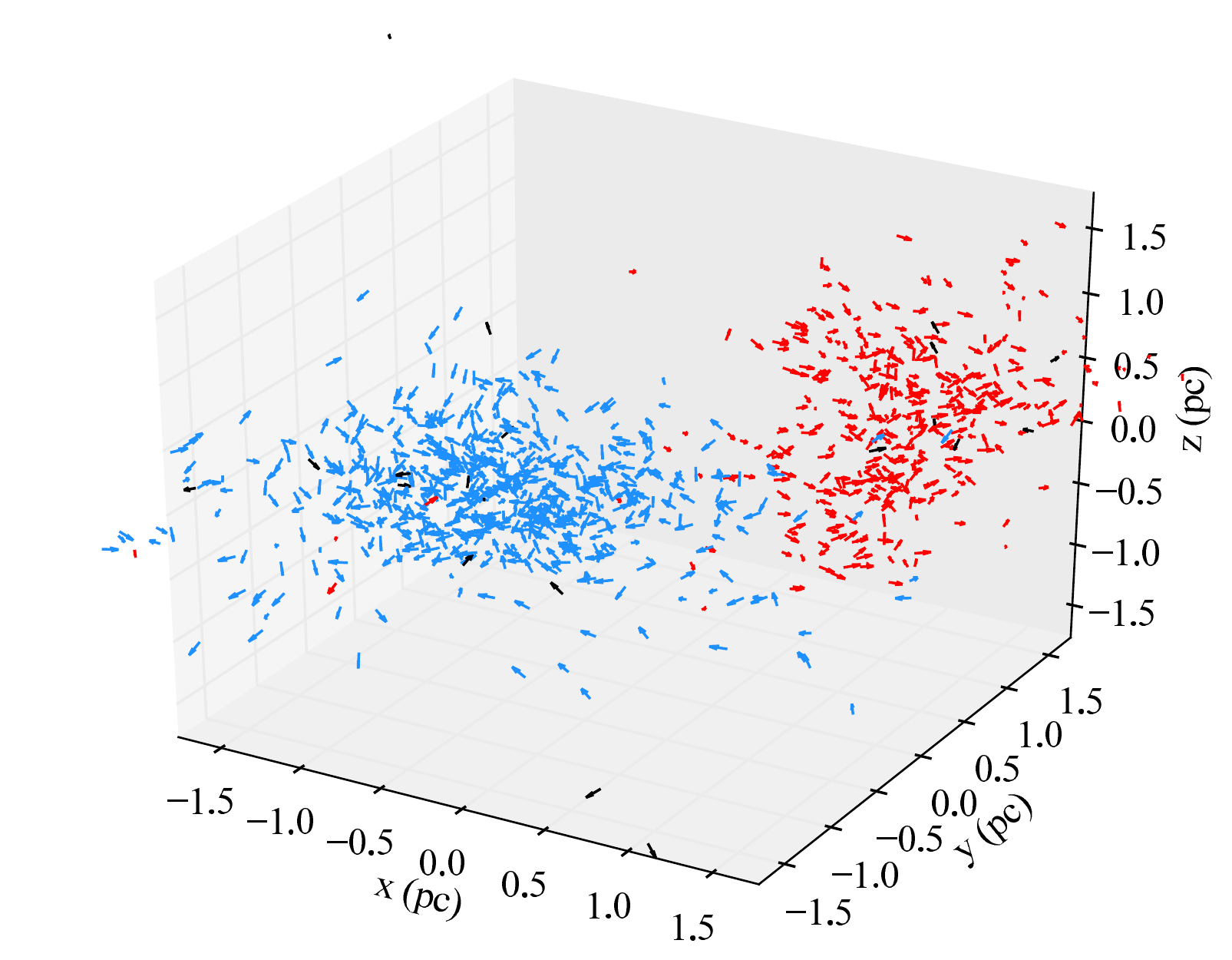}
    \caption{2.0 Myr}
    \label{2.0Myr}
  \end{subfigure}
  \caption{Snapshots at 0 Myr, 0.4 Myr, 0.8 Myr, 1.2 Myr, 1.6 Myr and 2 Myr of simulation (c) from Fig. 2.
    Stars are represented by arrows plotted in 3~pc-by-3~pc-by-3~pc boxes. The arrow's
    positions indicates star's positions in space, and the arrow's
    directions indicates the star's velocities. The arrows are colour
    coded: blue if the star is in the left hand cluster after the
    region finishes dividing, red if it is in the right cluster
    after division, and black means the star is unbound.}
  \label{sep_plot}
\end{figure*}

We run 50 simulations with the same input parameters as the previus set,
i.e. moderately-substructured and warm, but the velocities are randomised.
As one would expect all 50 of the regions evolve into
single clusters. This confirms that velocity structure is necessary for a binary
cluster to form.

Given the importance of velocity structure to the formation of binary
clusters, it is reasonable to wonder to what extent this might be an
artifact of the (somewhat unphysical) generation of velocity coherence via
inheritance.  To
test this we re-run the 12 simulations with coherence set up using a
different method:
velocities are sampled from a turbulent velocity field (see Section 2.2).
The initial spatial distributions of the simulations are unchanged.

Binary clusters form in two of these simulations
 (example shown in Fig.~4).   
In Fig.~5 we show the initial conditions of the realisation that evolves
into the binary cluster in Fig.~4: the blue stars
are those that end-up in the left-hand cluster, the red stars end-up
in the right-hand cluster, and black stars are unbound to either
cluster (cf. Fig.~3).

There is arguably less mixing in Fig.~5 than in the first
panel of Fig.~3; the blue points are mostly initially close together
towards the upper-centre. However, without colour coding it is still
not obvious that this part of the
initial conditions will produce a separate cluster.
So, as was the case in the simulations with inherited velocities,
 we argue that from the initial conditions (a) it is not at all
obvious whether a binary
cluster will be produced, and (b) it is impossible to say
which stars will end up in
which cluster (or be unbound).

To review, binary clusters form in 2/12 simulations which use a turbulent velocity
field, compared to 5/12 binary clusters from inherited
velocities. This test consists of too few simulations to estimate
the different rates of binary formation using each of the methods,
and a detailed investigation of different methods of setting-up
velocity coherence is beyond the
scope of this paper. The important points here are\\
1. Velocity coherence is necessary for binary clusters to form.\\
2. Two independent methods are used to generate coherence and binary formation
results from both. Therefore our results are not an artifact of the method used to initialise velocity structure.

\begin{figure}
  \includegraphics[width=\columnwidth]{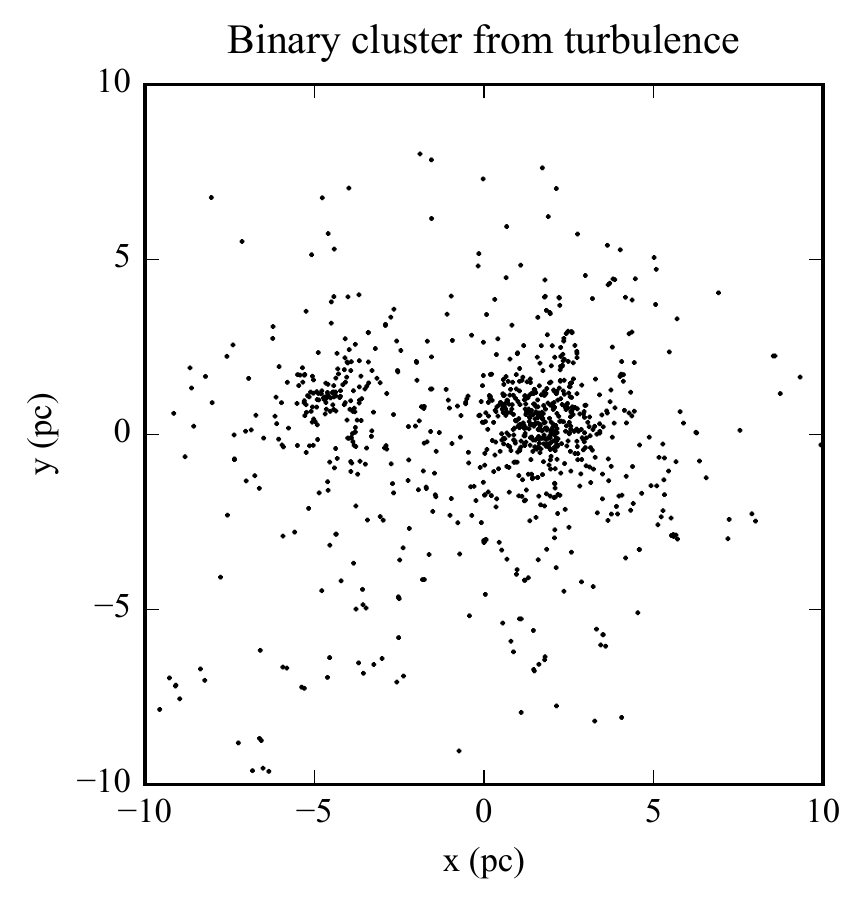}
  \caption{A 20~pc-by-20~pc $x$-$y$ projection of a simulation with initially
    turbulent velocities after 20 Myr. The region has developed into a binary cluster.}
  \label{turb_binary}
\end{figure}

\begin{figure*}
  \includegraphics[width=\textwidth]{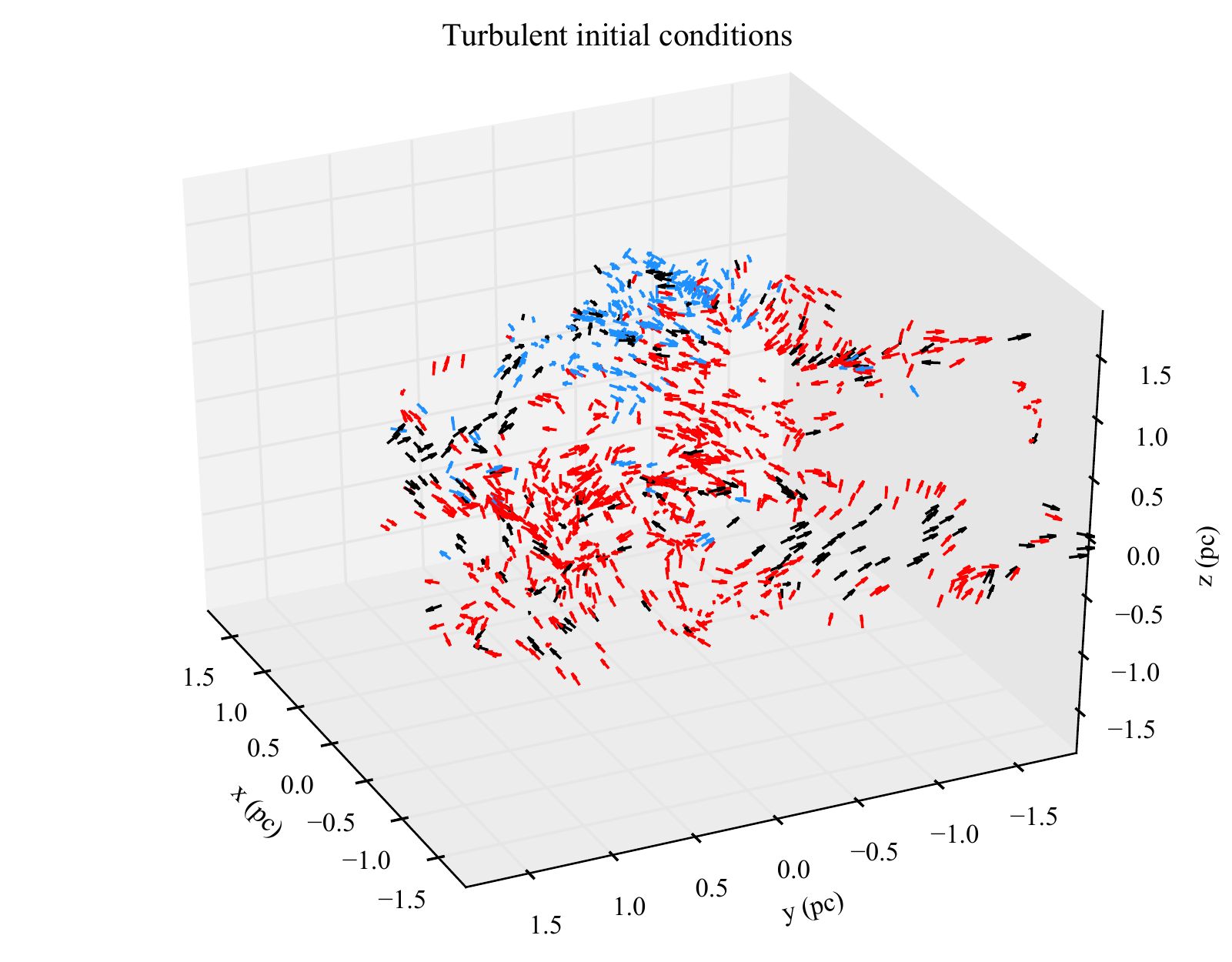}
  \caption{The initial conditions of the simulation shown in Fig. 4,
    which evolves into a binary cluster. Inspection of the figure clearly shows
    velocity coherence, which is produced by mapping the positions of the stars
    onto a turbulent velocity field. The colour coding is the same as
  described in Fig. 3.}
  \label{turb_init}
\end{figure*}

\section{Parameter space study}

In this section we explore parameter space to probe
which initial conditions can form binary clusters, and
investigate the properties of the binary clusters which form.  

As described in Section~2, nine ensembles of 50 simulations are performed.
The fractal dimension $D$ is varied such that $D$~=~$1.6$, 2.2, or 2.9
(highly-substructured (H), moderately-substructured (M), and smmoth (S)).
The virial factor $\alpha_{\rm vir}$ is varied such that $\alpha_{\rm vir}$~=~$0.3$, 0.5 or
0.7 (cool (C), virialised (V), or warm (W)). The simulations are run for 20 Myr, and are
summarised in Table~1.

\subsection{Which initial conditions produce binary clusters?}

We classify the final state of each simulation as one of three basic
categories.\\
{\bf Binary clusters:} two clearly distinguished clusters as
identified by the cluster finder and/or by eye. (In highly ambiguous
cases when the cluster finder struggles preference is given to the
by-eye conclusion).  Note that 4 of our 450 simulations
develop triple clusters. For the 
    sake of simplicity we classify these as binary clusters.\\
{\bf Single clusters:} one significant cluster (often with an unbound
`halo' of stars).\\
{\bf Binary merger:} a region that is a single cluster at 20~Myr,
but was a binary cluster at some earlier time. This may be because
a binary cluster formed and then merged into a single cluster, or
one of the two clusters dissolved.  Therefore,
depending on the time of an observation, they could be seen by an observer
as a (young) binary or a single cluster\footnote{The longest observed interlude between division
  and recombination of a binary merger in these simulations is $\sim 20$~Myr. At the other extreme, some binary mergers
  separate so briefly they are only `binary clusters' for $\sim 1$~Myr.}

The classifications of of the HV and HW simulations should be treated
with some caution as their long-lived substructure
makes several of them difficult to classify.
Four of the 450 simulations in this parameter space study
are deemed `unclassifiable', and are omitted.

In Fig.~6 we present the fractions of
regions which evolve into single, binary-merger
and binary clusters. These fractions approximate the probability
of each outcome, and the multinomial distribution is used to
calculated to one sigma confidence where the true probability lies,
which is indicated in Fig.~6 by error bars.

The top panel of Fig.~6 shows the results for the highly-substructured ($D$~=~$1.6$)
ensembles with virial ratios $\alpha_{\rm vir}$~=~$0.3$, 0.5, and 0.7 on the
$x$-axis.  The green circles are the fractions of single clusters,
yellow diamonds are the fraction of binary mergers,
and red diamonds the fraction of binary clusters.
The middle panel of Fig.~6 is the same plot but for the
moderately-substructured ($D$~=~$2.2$) ensembles, and the bottom panel is
for smooth ($D$~=~$2.9$) ensembles.

Each panel of Fig.~6 shows the same essential behaviour: binary
clusters are more common as the virial ratio increases.

When the regions are dynamically cool ($\alpha_{\rm vir}$~=~$0.3$, the left-most
results in each panel), almost all the simulations form a single
cluster.  This is as expected, as a dynamically cool distribution
will collapse and erase substructure (\citealt{Alison09b}; \citealt{Parker14}).
However, the cool ensembles also
produce some binary mergers;
even though these regions are collapsing, velocity structure
can allow them to `divide' for some amount of time.

When regions have moderate virial ratio ($\alpha_{\rm vir}$~=~$0.5$, the middle
results in each panel) the fraction of regions which evolve into
single clusters drops, and the fraction that evolve into binary
mergers increases concurrently.
The exception to
this is the H simulations, where both binaries and binary mergers
form.

When the regions are dynamically warm ($\alpha_{\rm vir}$~=~$0.7$, the right-most
results in each panel), the fraction of single clusters drops again and the
fraction of binary mergers drops somewhat at all levels of substructure.
In contrast, the fraction of binary clusters increases

The main result from ensembles of different initial conditions as
summarised in Fig.~6 are:\\
1) Higher $\alpha_{\rm vir}$ increases the probability that a region will divide\\
2) Binary clusters mainly form in dynamically warm regions.

\begin{figure}
  \setcounter{figure}{5}
         \begin{subfigure}[b]{\columnwidth}
                 \centering
                 \includegraphics[width=\columnwidth]{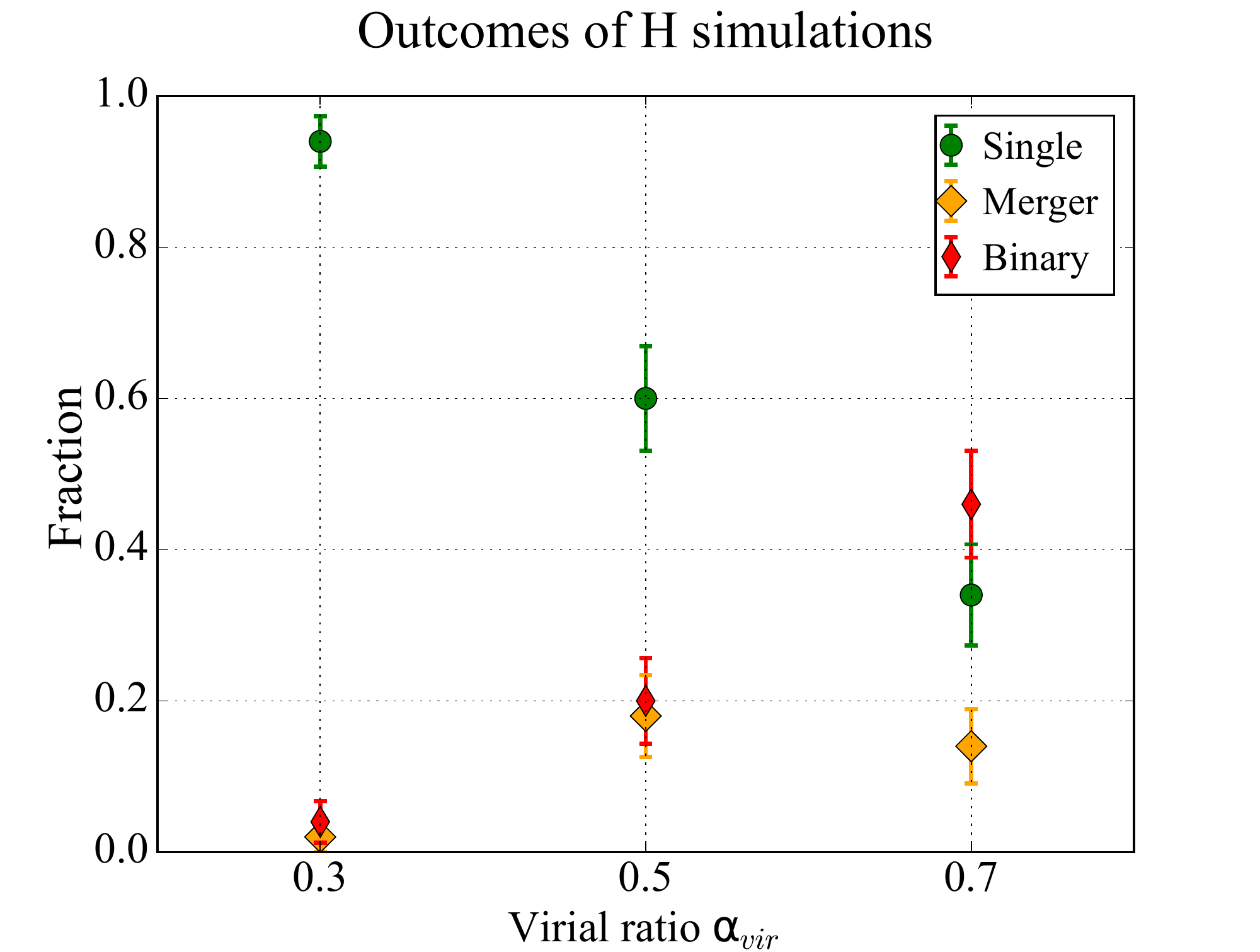}
                 \label{Outcome_H}
         \end{subfigure}

         \begin{subfigure}[b]{\columnwidth}
                 \centering
                 \includegraphics[width=\columnwidth]{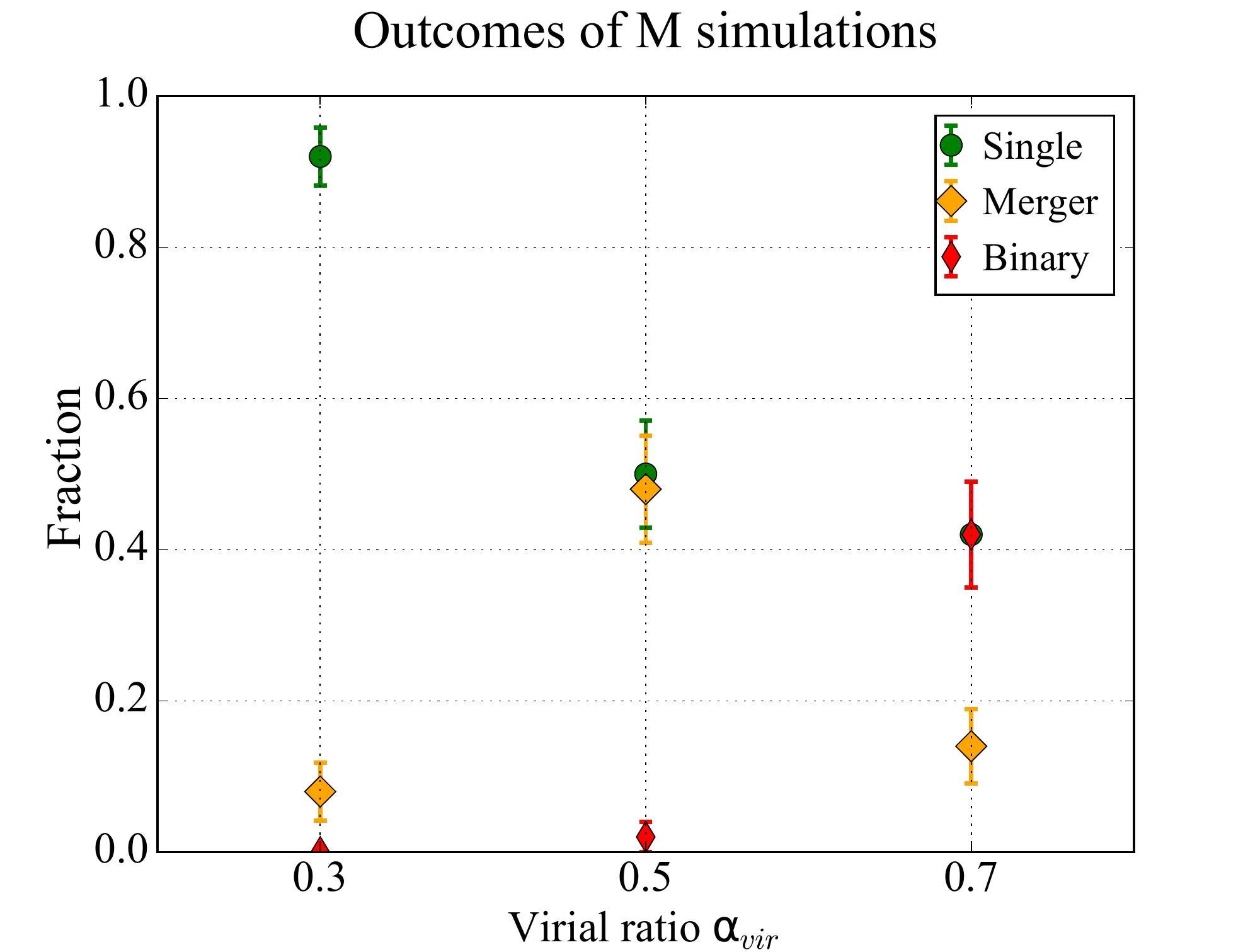}
                 \label{Outcome_M}
         \end{subfigure}

         \begin{subfigure}[b]{\columnwidth}
                 \centering
                 \includegraphics[width=\columnwidth]{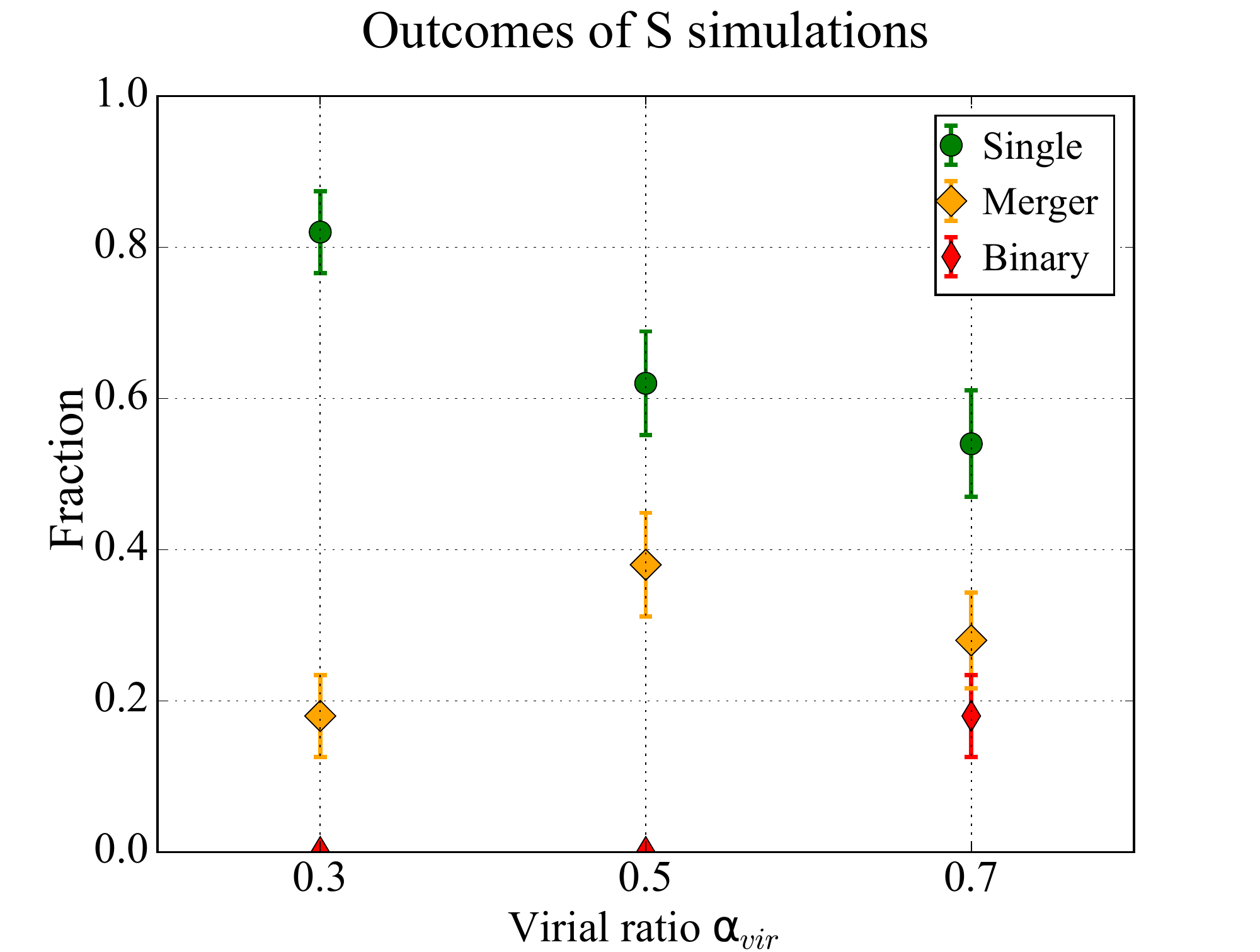}
                 \label{Outcome_S}
         \end{subfigure}
         \caption{The fraction of regions which evolve into single, binary-merger and binary clusters
           for each set of simulations.
           The highly-substructured (H) simulation results are shown
           in the top panel, the moderately-substructured (M) simulations
           in the middle panel, and the smooth (S) simulations in the bottom panel.
           The x axis separates the simulations by their virial ratio $\alpha_{\rm vir}$ (0.3, 0.5 or 0.7).
           The fraction of regions in a given set of
           simulations which evolve into single clusters is indicated
           by green circles. The fraction of binary-mergers
           is indicated by wide yellow diamonds, and the fraction of
           binary clusters is shown by narrow red diamonds.}
         \label{Outcome_plot}
\end{figure} 

\subsection{`Micro-clusters'}

Regions do not always divide into a clean binary or single clusters.
In particular, the highly-substructured regions
($D$~=~$1.6$), especially with high virial ratio ($\alpha_{\rm vir}$~=~$0.5$, 0.7),
can often form several small, bound objects we
refer to as `micro-clusters'.  Fig.~7 shows an example of an HW
simulation at 20~Myr with four micro-clusters (indicated by the red
arrows).  

Whilst these micro-clusters are able to survive 20~Myr, they will have
short lifetimes because they only contains tens of members
so their two-body relaxation time is very short. Nevertheless, they could be
observed around young clusters and mistaken for independent objects
instead of potential evidence
that the region was initially highly substructured.

\begin{figure}
  \includegraphics[width=\columnwidth]{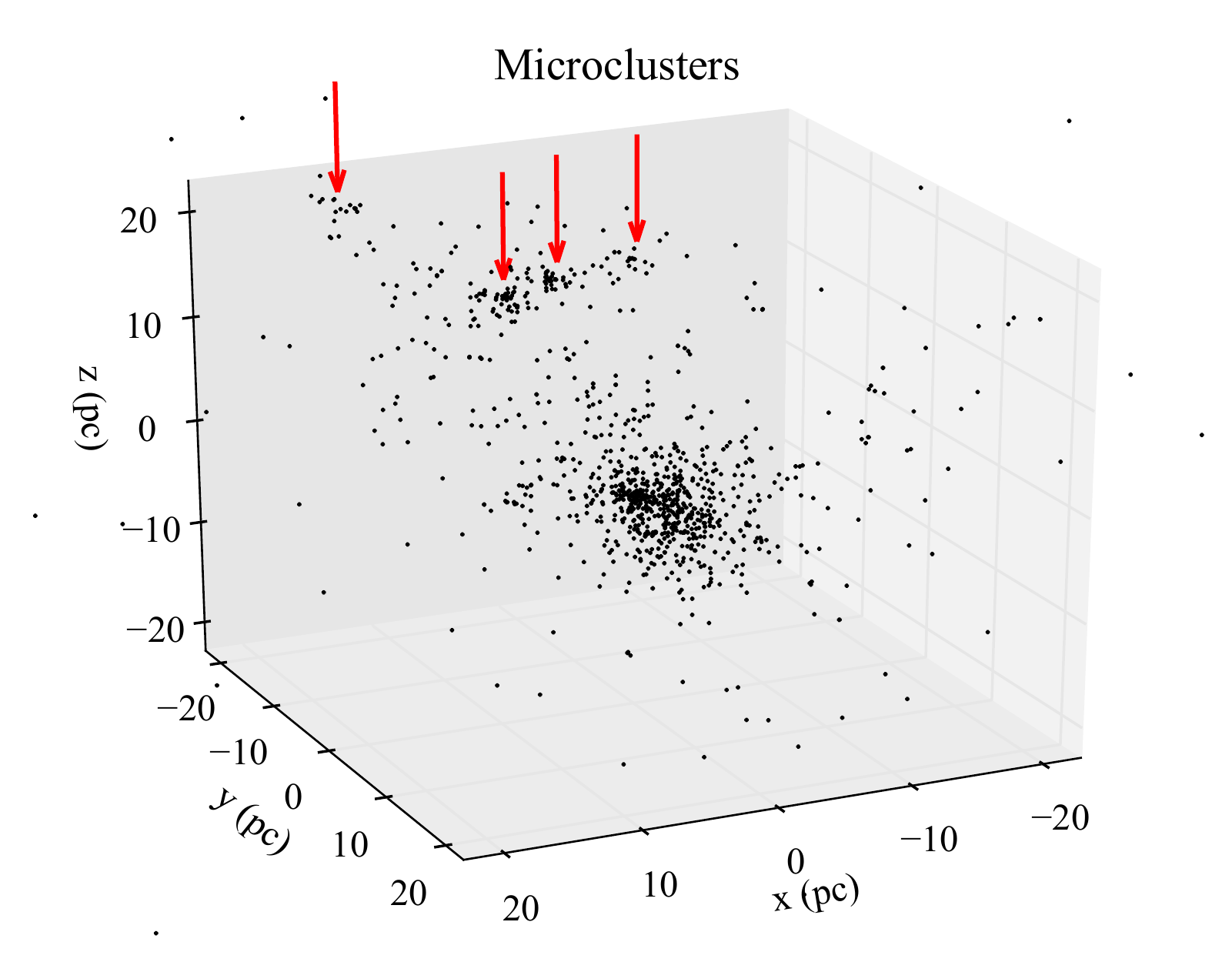}
  \caption{A highly-substructured, dynamically warm region that
    has been evolved for 20 Myr. The star's positions are
    indicated by dots in a 40~pc-by-40~pc-by-40~pc box.
    The simulation has developed numerous long lived overdensities.
    We call these overdensities `microclusters', and they are highlighted by red arrows.}
  \label{microclusters}
\end{figure}

\subsection{Elongated clusters}

Some S regions undergo a period of elongation
followed by collapse within the first $\sim$5 Myr.
This is observed mainly in S simulations, and
is most common when $\alpha_{vir}$ is high.
An example is shown in Fig.~8

\begin{figure}
  \includegraphics[width=\columnwidth]{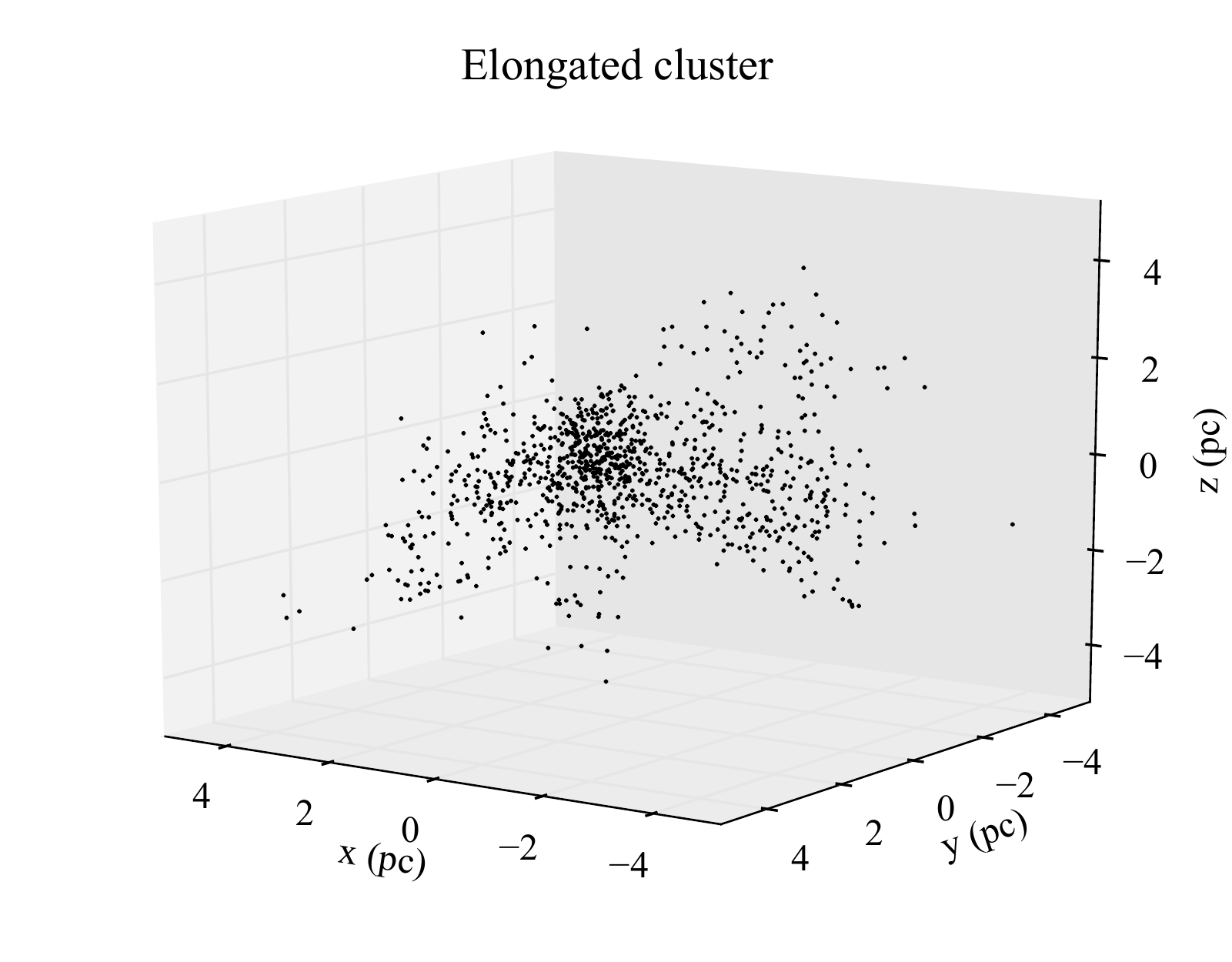}
  \caption{An initially smooth, virialised region after
    evolving for 3 Myr. It has developed an elongated shape
    which collapses back
  into a sphere within the next few Myr.}
  \label{elongated}
\end{figure}

\subsection{Cluster mass ratios}

We determine the mass ratios of the binary clusters,
(and binary-mergers when they are distinct entities)
using the cluster finding algorithm. In highly ambiguous cases,
where the algorithm struggles, the cluster memberships are determined by eye.
This means that in some cases a {\em  particular} mass ratio should be treated with caution.
However the trends we describe should not
be affected by a small number of ambiguities. Note that we define mass
ratios as the mass of the lighter cluster divided by the mass of the heavier one. Therefore if the cluster masses are very different the mass ratio is low,
and if their masses are fairly equal the mass ratio is high,
with a maximum of unity.

The cumulative distribution functions (CDFs) of highly-substructred
(H, black line), moderately-substructed (M, purple line) and smooth
(S, red line) binary mass ratios are shown in Fig.~9.  Most binary
clusters are from warm simulations ($\alpha_{\rm vir}$~=~$0.7$) as these are
the ensembles that produce the vast majority of binary clusters.

From Fig.~9 it is clear that the mass ratio distributions
for each level of substructure are distinct (a KS test gives a
$P$-value $< 10^{-4}$ for any pair of distributions, confirming that they
are statistically different).

Binary clusters that form from highly-substructred initial conditions
(H, black line) tend to have low mass ratios (i.e very unequal
cluster masses) almost all being 0.1--0.4
(median 0.3).

Binary clusters that form from smooth initial conditions (S, red line)
have higher mass ratios, almost all between 0.3--0.6 (median 0.44).

Binary clusters that form from moderately-substructred initial conditions 
(M, purple line) have generally higher mass ratios still, ranging mostly
from 0.4--0.8 (median 0.54).

We may have expected to see a sequence in mass ratio
distributions that moves from highly-substructred, to
moderately-substructred, to smooth. Instead the
smooth region's mass ratios are intermediate between those of the highly
and moderately-substrucured regions.

We explain this by first considering the smooth regions (red line)
as a baseline.
They have no spatial structure, so their mass ratios are entirely due to
velocity structure.

The moderately-substructured regions contain fairly large spatial
structures, which themselves are correlated with the velocity structure.
As a result there are often natural `starting points' for sizable
portions of the regions to separate from the rest, resulting in more even
mass ratios.

In contrast, the highly-substructured regions contain many small spatial
structures which may divide from the main cluster, resulting in many
low mass ratio systems.

\begin{figure}
  \includegraphics[width=\columnwidth]{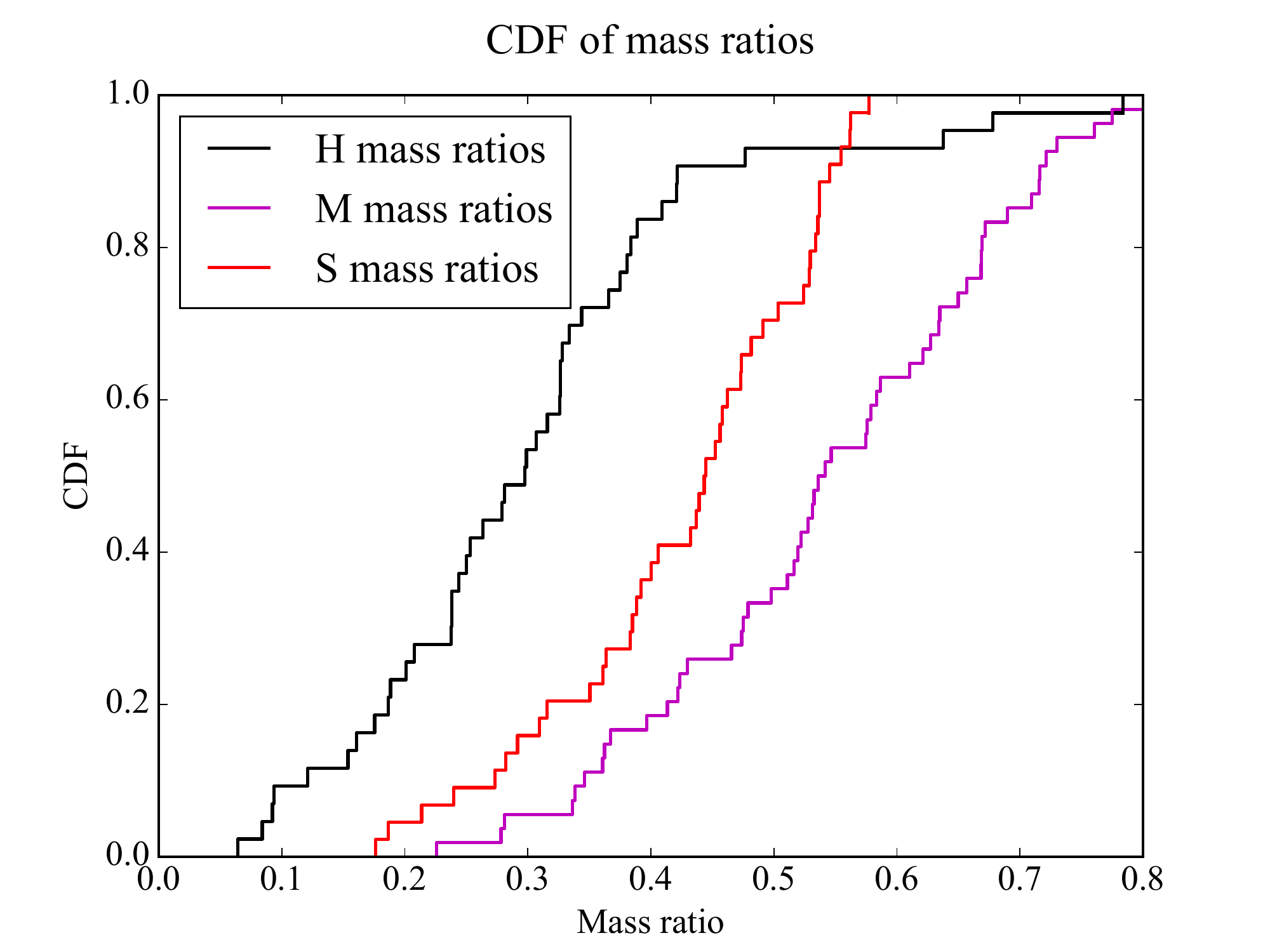}
  \caption{Cumulative distribution functions of the cluster mass ratios.
    The CDF for highly-substructured (H)
    simulations is plotted by a black line, the moderately-substructured (M) simulations by a magenta line and the smooth (S) simulations
    by a red line.}
  \label{mass_ratios}
\end{figure}

\subsection{Comparison with observations}

Probably the best-known binary clusters, h and $\chi$ Per, have a
mass ratio of 0.76 (masses of 3700 and 2800 $M_\odot$ respectively,
\citealt{Slesnick02}), which, from Fig.~9,
suggests moderate initial substructure.
However this is only a single binary
cluster, which may not be representative of the
conditons that the majority binary clusters form from.

A catalogue of well measured mass ratios
could potentially provide observational clues as to the
initial conditions of the regions that produce real-world binary clusters. 
Unfortunately, the current state of observational data cannot
provide good constraints \citep{Conrad17}.
However, if more than one mechanism is responsible for
  binary cluster formation, the cluster mass ratio distribution would be a combination of those produced by all the different mechanisms, and would be much harder to interpret.

\section{Conclusions}

We perform ensembles of $N$-body simulations of $N$~=~$1000$, $R$~=~$2$~pc
regions, which are evolved for 20 Myr.  These regions
start with fractal dimensions of $D$~=~$1.6$, 2.0 or 2.9 (from
highly-substructured to smooth), and 
virial ratios (the ratio of
kinetic to potential energies) of $\alpha_{\rm vir}$~=~$0.3$, 0.5 or 0.7 (from cool to warm).
The velocities of stars are `coherent'; stars that are initially close
together tend to have similar velocities.

We find that single star forming regions can dynamically
  evolve into binary clusters, 
(although this is not necessarily the only way binary clusters may 
form).
We find that initial velocity structure is necessary for a region to divide,
and in all cases it is essentially impossible to determine 
from the initial state
of a region:\\
1. If a binary cluster will form (although most of the regions that {\em  do} form binaries 
are initially dynamically warm ($\alpha_{\rm vir}$~=~$0.7$)).\\
2. Which stars will end up in
which component of the binary cluster.

The two clusters move directly apart from one another with relative velocities
typically $\sim 1$~km~s$^{-1}$, so pairs will appear associated for 10s~Myr.
In some cases the clusters remain bound to one another, and recombine at a
later time. We describe these as binary-mergers, and they are most common
in regions that begin in virial equilibrium. 

We find that the level of initial spatial structure
in a region strongly influences the
mass ratio of a resulting binary cluster.
High levels of
initial substructure tend to result in very un-equal masses (mass
ratios 0.2--0.4), no initial spatial substructure results in slightly more
equal mass ratios (0.3--0.6), and moderate substructure in even more
equal mass ratios (0.4--0.8).

\section*{Acknowledgements}

BA acknowledges PhD funding from the University of 
Sheffield, and DWG from the STFC. RJP acknowledges support from
the Royal Society in the form of a Dorothy Hodgkin Fellowship.
Thanks also to S{\o}ren Larsen for useful discussions.




\bibliographystyle{mnras}
\bibliography{mybibfile}


\section*{Appendix A: The cluster-finding algorithm}

We briefly describe how clusters are identified in a snapshot of the simulation.

\begin{itemize}
\item Step 1: Distinguish areas of high stellar density.
  
  Space is divided into equally sized boxes by a three dimensional grid. The resolution of this grid is initially low, and it is increased until 75 per cent of the stars of the stars are contained within at least 20 boxes. This resolution distinguishes areas of high stellar density without being too fine or coarse.
  
\item Step 2: Find the position of a cluster.
  
  The box containing the most stars is located. By definition, clusters are regions with many stars so this box will be at, or close to, the centre of a cluster. The centre of mass and centre of velocity of the stars in this box is calculated. The mass of stars in this box is used to crudely estimate the mass of the cluster.
  
\item Step 3: Identify cluster members.
  
  The program goes through each star and calculates its kinetic and potential energy relative to the position, velocity, and mass determined in step 2. Bound stars are identified as cluster members.
  
\item Step 4: Identify further cluster members.
  
  The centre of mass, centre of velocity, and total mass of the cluster members is calculated. Step 3 is repeated using these values, i.e. the kinetic and potential energy of each star relative to this position, velocity, and mass is calculated to identify further members.
  
\item Step 5: Find additional clusters.

  Steps 2-5 repeat. In order to prevent the same cluster being identified multiple times, stars that have already been identified as members of a cluster are excluded in step 2. Therefore when all the clusters have been identified the box containing the most stars, as identified by step 2, contains so few stars it could not reasonably be the centre of a new cluster. The program stops searching for additional clusters after that point. Any remaining stars are determined to be unbound.

\item Step 6: Clean up.

  This step prevents the order in which the clusters are identified from
    influencing the final membership lists.
    All information on which stars belong to which cluster is thrown away; only the positions, masses, and velocities of the clusters are retained. The potential and kinetic energy of each star compared to these clusters is calculated to determine which cluster (if any) the star is most strongly bound to. This produces the final membership list for each cluster. The mass, centre of mass, and centre of velocity of each cluster is recalculated using this membership list.
  
\end{itemize}

\bsp	
\label{lastpage}
\end{document}